\documentclass[twocolumn]{aastex631}

\newcommand\xmm{\textit{XMM-Newton}}
\newcommand\chandra{\textit{Chandra}}
\newcommand\maxi{\textit{MAXI}}
\newcommand\bat{\textit{Swift/BAT}}
\newcommand\swift{\textit{Swift}}
\newcommand\nustar{\textit{NuSTAR}}
\newcommand\suzaku{\textit{Suzaku}}
\newcommand\rxte{\textit{RXTE}}

\newcommand\delcstat{$\Delta$C-stat}
\newcommand\pcm{cm$^{-2}$}
\newcommand\RG{R$_{\rm{G}}$}

\shorttitle{The long stare at Her X-1 I. Emission lines}
\shortauthors{Kosec et al.}
\graphicspath{{./}{figures/}}

\begin{document}

\title{The long stare at Hercules X-1 - I. Emission lines from the outer disk, \\ the magnetosphere boundary and the accretion curtain}

\correspondingauthor{P. Kosec}
\email{pkosec@mit.edu}

\author{P. Kosec}
\affiliation{MIT Kavli Institute for Astrophysics and Space Research, Massachusetts Institute of Technology, Cambridge, MA 02139}

\author{E. Kara}
\affiliation{MIT Kavli Institute for Astrophysics and Space Research, Massachusetts Institute of Technology, Cambridge, MA 02139}

\author{A. C. Fabian}
\affiliation{Institute of Astronomy, Madingley Road, CB3 0HA Cambridge, UK}

\author{F. Fürst}
\affiliation{Quasar Science Resources SL for ESA, European Space Astronomy Centre (ESAC), Science Operations Departement, 28692 Villanueva de la Can\~nada, Madrid, Spain}

\author{C. Pinto}
\affiliation{INAF - IASF Palermo, Via U. La Malfa 153, I-90146 Palermo, Italy}

\author{I. Psaradaki}
\affiliation{University of Michigan, Dept. of Astronomy, 1085 S University Ave, Ann Arbor, MI 48109, USA}

\author{C. S. Reynolds}
\affiliation{Institute of Astronomy, Madingley Road, CB3 0HA Cambridge, UK}

\author{D. Rogantini}
\affiliation{MIT Kavli Institute for Astrophysics and Space Research, Massachusetts Institute of Technology, Cambridge, MA 02139}

\author{D. J. Walton}
\affiliation{Centre for Astrophysics Research, University of Hertfordshire, College Lane, Hatfield AL10 9AB, UK}

\author{R. Ballhausen}
\affiliation{Department of Astronomy, University of Maryland, College Park, MD 20742, USA}
\affiliation{NASA-GSFC/CRESST, Astrophysics Science Division, Greenbelt, MD 20771, USA}

\author{C. Canizares}
\affiliation{MIT Kavli Institute for Astrophysics and Space Research, Massachusetts Institute of Technology, Cambridge, MA 02139}

\author{S. Dyda}
\affiliation{Department of Astronomy, University of Virginia, 530 McCormick Rd, Charlottesville, VA 22904, USA}

\author{R. Staubert}
\affiliation{Institut für Astronomie und Astrophysik, Universität Tübingen, Sand 1, 72076 Tübingen, Germany}

\author{J. Wilms}
\affiliation{Dr. Karl Remeis-Observatory and Erlangen Centre for Astroparticle Physics, Sternwartstr. 7, D-96049 Bamberg, Germany}



\begin{abstract}

Hercules X-1 is a nearly edge-on accreting X-ray pulsar with a warped accretion disk, precessing with a period of about 35 days. The disk precession allows for unique and changing sightlines towards the X-ray source. To investigate the accretion flow at a variety of sightlines, we obtained a large observational campaign on Her X-1 with \xmm\ (380 ks exposure) and \chandra\ (50 ks exposure) for a significant fraction of a single disk precession cycle, resulting in one of the best datasets taken to date on a neutron star X-ray binary. Here we present the spectral analysis of the \textit{High State} high-resolution grating and CCD datasets, including the extensive archival data available for this famous system. The observations reveal a complex Fe K region structure, with three emission line components of different velocity widths. Similarly, the high-resolution soft X-ray spectra reveal a number of emission lines of various widths. We correct for the uncertain gain of the EPIC-pn Timing mode spectra, and track the evolution of these spectral components with Her X-1 precession phase and observed luminosity. We find evidence for three groups of emission lines: one originates in the outer accretion disk ($10^5$ \RG\ from the neutron star). The second line group plausibly originates at the boundary between the inner disk and the pulsar magnetosphere ($10^3$ \RG). The last group is too broad to arise in the magnetically-truncated disk and instead must originate very close to the neutron star surface, likely from X-ray reflection from the accretion curtain ($\sim10^2$ \RG).

\end{abstract}



\section{Introduction} \label{sec:intro}

Hercules X-1 (hereafter Her X-1) is a famous neutron star X-ray binary \citep{Giacconi+72, Tananbaum+72} widely known for the various timescales of its variability. In addition to the 1.24 s rotation period of the neutron star \citep{Giacconi+73, Deeter+81, Staubert+09} and the 1.7 d binary period \citep{Davidsen+72, Bahcall+72, Middleditch+76}, the system exhibits a strong flux variability with a period of about 35 days \citep{Katz+76, Gerend+76}, the so-called super-orbital cycle. Each cycle begins with a \textit{Turn-on} during which the object switches into an X-ray bright \textit{High State} with a luminosity of $\sim3 \times 10^{37}$ erg/s for $7-10$ days. This is followed by a \textit{Low State} with an X-ray flux of about $5-10$\% of the \textit{High State}. Another, weaker high flux state (with a flux of about 1/3 of the \textit{High State} flux), called the \textit{Short High} \citep{Fabian+73} then occurs for roughly a week, and finally a second \textit{Low State} follows until the end of each super-orbital cycle. The cycle has been studied since 1970s and is not strictly periodic and varies in length between 33 and 37 days \citep{Leahy+10}.

These variations have been studied in great detail and can be explained within a scenario where the accretion disk of Her X-1 is seen almost edge on, is warped and precesses each super-orbital period \citep{Roberts+74, Peterson+75, Scott+00, Ogilvie+01}. In such a case, the compact X-ray source is obscured from our view during the \textit{Low States} by the outer accretion disk, while directly observable during the \textit{High States}.


Her X-1 exhibits a great variety of phenomena and its spectra are highly complex. It has been studied thoroughly thanks to its proximity \citep[6.1 kpc,][]{Leahy+14}, combined with regularly appearing high X-ray flux and thus provides us the opportunity to obtain very high quality X-ray spectra with current observatories such as \xmm\ and \chandra. Despite the apparent changes in flux due to the obscuration by the precessing disk, Her X-1 is stably accreting with a mass accretion rate of about 20\% of the Eddington limit for a canonical 1.4 M$_{\odot}$ neutron star. The broadband spectrum is composed of a pulsating Comptonization continuum from the accretion column of the neutron star \citep{Lamb+73, Ghosh+79}, as well as a soft blackbody originating from the reprocessed primary component \citep{Hickox+04}. A cyclotron resonance scattering feature modifies the spectrum around 40 keV \citep{Truemper+78, Staubert+07}, providing information on the magnetic field of the neutron star ($\sim 3 \times10^{12}$ G).

The X-ray spectrum also contains many emission lines of varying widths \citep{Pravdo+77, Ramsay+02, Jimenez+02, Jimenez+05}, from narrow lines ($\sim$100 km/s velocity widths) to strongly broadened features ($>$10000 km/s widths), plausibly originating through a number of different phenomena from distinct regions of the accretion flow.

Despite the number of previous studies, Her X-1 still has not revealed all its secrets. Recently, \citet{Kosec+20} detected a highly ionized and highly variable accretion disk wind through narrow absorption lines seen in \xmm\ spectra. The existence of an outflow in Her X-1 has previously been predicted by \citet{Schandl+94} and \citet{Leahy+15}. We proposed that the observed strong wind parameter variation may be explained if the different observations are sampling differing sightlines above the precessing warped disk. Therefore the sightlines are sampling different parts (heights) of the disk wind structure at different phases of the super-orbital precession cycle. This discovery thus revealed a unique way to study the 3D structure of an accretion disk wind in an X-ray binary.

\begin{figure*}
\includegraphics[width=\textwidth]{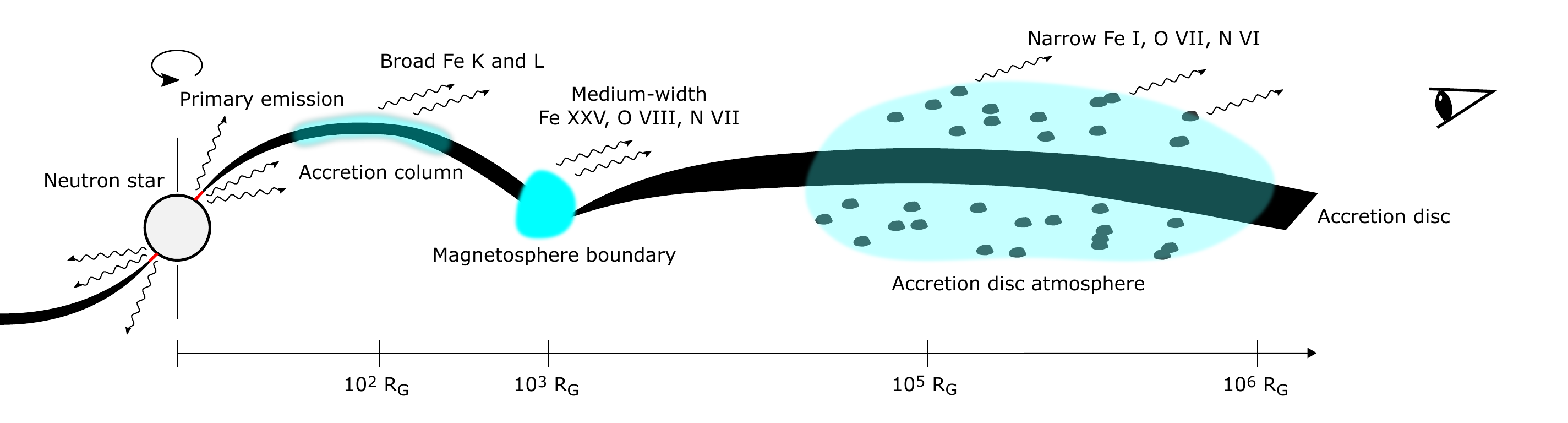}
\caption{A schematic of the Her X-1 system, showing the different regions producing emission lines of various velocity widths. The X-axis of the schematic is in logarithmic scale. For simplicity, the scheme omits the accretion disk wind (detected through its absorption lines), likely originating from the inner parts of the accretion disk atmosphere. \label{HerX1_scheme}}
\end{figure*}

To follow-up on these results, we were awarded a Large (380 ks) \xmm\ campaign on Her X-1 in AO-19 (PI: Kosec). The primary aim was to study the evolution of the disk wind over the course of a single \textit{High State}. At the same time, the campaign produced a great wealth of serendipitous data on Her X-1. In this paper, we focus on the properties of the numerous and diverse emission lines found in its X-ray spectrum, identifying them with three distinct regions of the accretion flow: the outer accretion disk, the boundary between the disk and the neutron star magnetosphere and the accretion curtain (or column) of the pulsar. Fig. \ref{HerX1_scheme} contains a schematic of the Her X-1 system and shows our interpretation of the origin of these emission lines. The analysis of the accretion disk wind properties will be presented in a separate forthcoming publication (Paper II).

The structure of this paper is as follows. In section \ref{sec:lightcurve}, we describe the new August 2020 campaign and show the \xmm\ Reflection Grating Spectrometer (RGS) lightcurve of Her X-1. Section \ref{sec:dataprep} contains the details of our data preparation and reduction, while in section \ref{sec:modelling} we describe the spectral model used to fit the Her X-1 spectra from \chandra\ and \xmm\ instruments. In section \ref{sec:results} we present the results of this study and in section \ref{sec:discussion} we discuss their implications. We summarize in section \ref{sec:conclusions}. Finally, appendices \ref{app:pngain} and \ref{app:ignrgs} contain further technical details about spectral modelling.

Throughout the paper, we adopt a distance of Her X-1 of 6.1 kpc \citep{Leahy+14}. All uncertainties are provided at 1$\sigma$ significance, unless explicitly specified otherwise.

\section{August 2020 \xmm\ and \chandra\ Campaign on Her X-1} \label{sec:lightcurve}

We were awarded a 380 ks \xmm\ exposure to study the evolution of the highly variable ionized disk wind in Her X-1. The campaign was composed of three back-to-back full-orbit observations to track the wind variation during a significant part of a single \textit{High State} of Her X-1. The aim was to catch the \textit{Turn-on} moment of the \textit{High State}, and observe for the next $\sim400$ ks, interrupted only by \xmm's orbit. Two additional \chandra\ observations (35 ks + 15 ks) were obtained through a DDT request to provide coverage during the first \xmm\ observation gap, to resolve better the Fe K energy band as well as to compare the results from the two observatories for consistency. 

\begin{figure*}
\includegraphics[width=\textwidth]{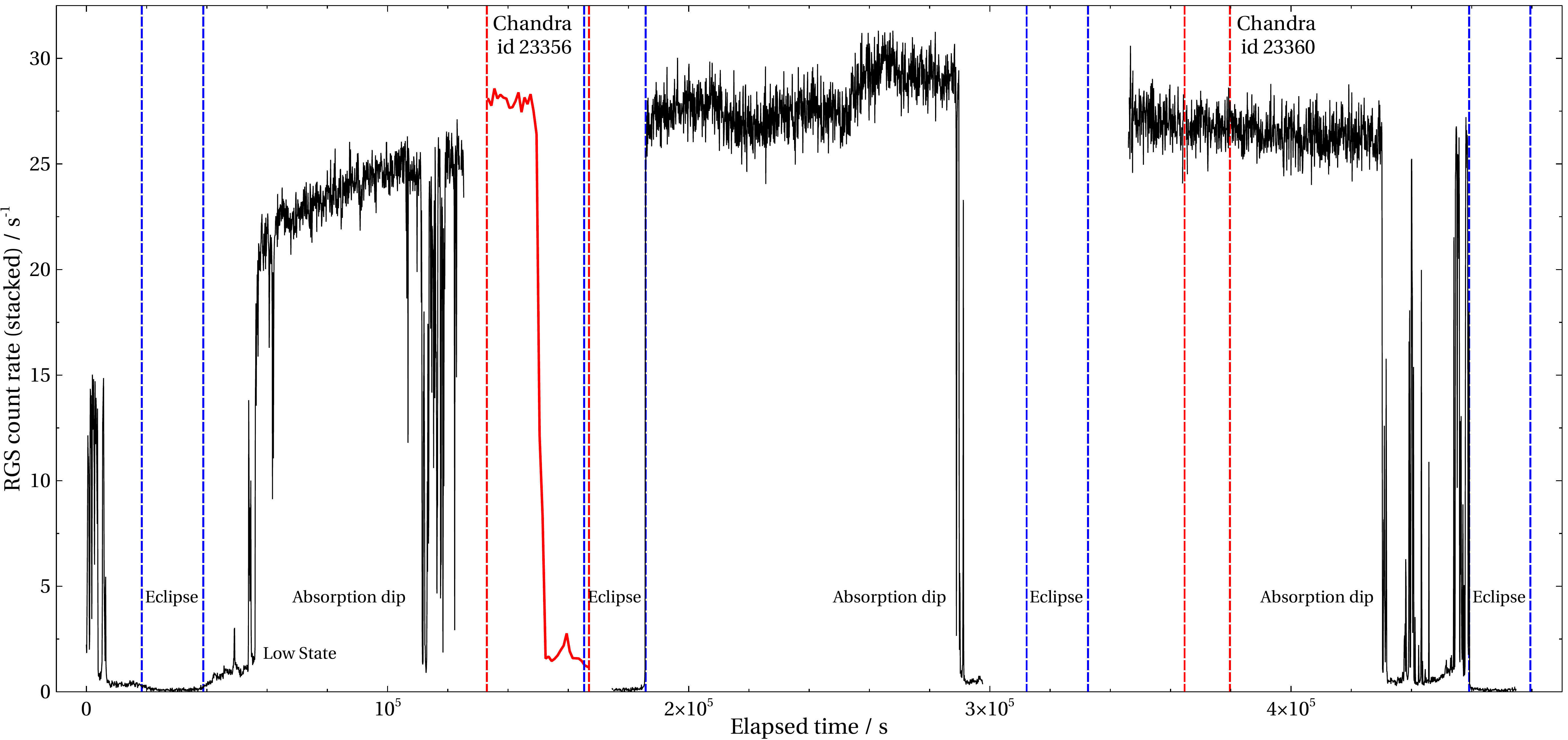}
\caption{\xmm/RGS (stacked) $0.35-1.8$ keV and \chandra\ HETG ($1-9$ keV) lightcurves from the August 2020 campaign on Her X-1, with a resolution of 100 s. The campaign consisted of three full-orbit \xmm\ observations as well as two shorter \chandra\ observations. Aside from the \textit{High State} and a brief low flux period, we observed one full and two partial eclipses of Her X-1 by the secondary and four absorption dipping periods. The $0.3-10$ keV EPIC pn lightcurve has a very similar shape to the RGS lightcurve, but the absorption dips are less pronounced due to the higher energy coverage of pn. \chandra\ count rate during observation 23356 was scaled by the RGS/HETG count ratio during observation 23360 with simultaneous coverage of both instruments. The lightcurve of \chandra\ observation 23360 is not shown as the same time interval is covered by \xmm. The elapsed time of 0 corresponds to MJD 59071.69035. \label{xcamp_lcurve}}
\end{figure*}

Our observations were triggered and carried out between the 10th and the 16th August 2020. The RGS lightcurve of the three \xmm\ observations is shown in Fig. \ref{xcamp_lcurve}. At the very beginning of the first observation, Her X-1 was already in the \textit{High State}, with an RGS count rate of $10-15$ ct/s ($30-50$ \% of the maximum during this \textit{High State}). We inspected the concurrent \bat\ and \maxi\ monitor lightcurves to determine the time of \textit{Turn-on} (here defined as the first moment when Her X-1 shows high flux after the second \textit{Low State} of the previous precession cycle). We found that the \textit{Turn-on} could not have occurred more than a few ks before the \xmm\ campaign began, and estimated the Turn-on moment to have occurred at MJD 59071.65. The initial brief high flux state was followed by a low flux period including an eclipse of the X-ray source by the secondary HZ Her. Then, around halfway through the first \xmm\ observation, the \textit{High State} returned for the rest of the campaign. 

The observed X-ray spike into high X-ray flux, followed by a low flux period before the persistent \textit{High State} continues is an interesting phenomenon, which was not seen before in such clarity. It is most likely connected with the outer edge of the accretion disk not being a sharp edge but a highly structured entity of scattering and absorbing material. A similar highly variable light curve was observed by \xmm\ in August 2016. This complex \textit{Turn-on} behavior will be studied in a separate publication.

In addition to the high flux periods, the campaign covered one full and two partial eclipses of Her X-1 by the secondary HZ Her, as well as three dipping periods. The flux dips originate from cold matter absorption in the outer accretion disk, and are possibly connected to the impact of the accretion stream onto the disk \citep{Choi+94, Schandl+95}. The dips and eclipses will be described and studied in detail in forthcoming publications.

The extreme Her X-1 count rate and the very long exposure result in one of the best quality \xmm\ datasets on a neutron star X-ray binary, captured in high spectral resolution. As the goal of the campaign was to track the time evolution of the accretion disk wind properties, we decided to split the full campaign in a number of smaller segments. We split the campaign into 14 \textit{High State} segments in total, achieving a compromise between time sampling and data quality. The segmenting is described in section \ref{sec:dataprep}.


\section{Data Preparation and Reduction} \label{sec:dataprep}

In this work we analyze spectra from \xmm\ and \chandra\ observations. We use all sufficient quality archival and new observations of the \textit{High State} of Her X-1 with these instruments. The backbone of the work is the Large \xmm\ campaign which occurred in August 2020 (380 ks raw exposure), but the \xmm\ Her X-1 archive contains 11 further \textit{High State} observations (190 ks raw exposure in total). The August 2020 campaign was accompanied by two \chandra\ DDT observations (PI: Kosec) with 50ks of exposure in total. We also analyze an additional 20ks archival \chandra\ observation. 

The new \xmm\ observations are split into smaller segments in order to perform time-resolved spectroscopy. We calculate the super-orbital (precession) phase for each observation or observation segment analyzed. In this work we consider the \textit{Turn-on} moment (phase=0) to be the first moment when the apparent luminosity of Her X-1 spikes to its \textit{High State} value (roughly 10$^{37}$ erg/s and higher, about $\sim$25 \% of the maximum observed \textit{High State} luminosity). We use \bat\ \citep{Krimm+13} and \maxi\ \citep{Matsuoka+09} lightcurves to determine the \textit{Turn-on} moments for the current and the following precession cycle, and determine the super-orbital phase, following the approach in \citet{Kosec+20}. We estimate the uncertainties on each phase measurement. These are significantly smaller for the August 2020 observational campaign where the \textit{Turn-on} moment can be very well constrained. The super-orbital phases of observations 0134120101 and 0153950301 (determined from \rxte\ data) were obtained from \citet{Leahy+10}.

We also determine the orbital phase of each observation/segment. We take the midpoint of the clean exposure and calculate the orbital phase and its uncertainty (taken to be half the segment duration) using the Her X-1 orbital solution from \citet{Staubert+09}.


\subsection{\xmm}


\xmm\ \citep{Jansen+01} observed Her X-1 many times in the past. However, many of the exposures occurred when the object was in its \textit{Low State} and are not analyzed here. There are 11 archival \textit{High State} observations (typically of 10-20 ks duration), most of them previously also analyzed by \citet{Kosec+20}. The new \textit{High State} observations are split into 14 segments of varying exposure. The splits were chosen manually and were motivated by the strength of the ionized wind absorption features, the strength of which reduces with increasing super-orbital phase. The first observation, 0865440101, is split into 7 segments. The second observation, 0865440401, is split into 5 segments, while the final observation, 0865440501, into 2 segments. The details of the archival observations are shown in Table \ref{xarchdata} and the details of the new observations (split into 14 segments) are in Table \ref{xcampdata}.

All \xmm\ data were downloaded from the XSA archive and reduced using a standard pipeline with \textsc{sas} v19, \textsc{caldb} as of 2021 June. We analyze data from the Reflection Grating Spectrometers \citep[RGS, ][]{denHerder+01} and from the European Photon Imaging Camera (EPIC) pn instrument \citep{Struder+01}. We do not use Optical Monitor data due to confusion with the secondary HZ Her, which is also bright in the optical and near-UV bands, and time variable due to the illumination from the inner accretion flow.

\begin{deluxetable*}{ccccccccccc}
\tablecaption{Details of archival \xmm\ observations used in this work. All exposures listed are clean. The pn count rates shown are after pile-up correction. All Turn-on MJD dates have an uncertainty of 0.5 day. \label{xarchdata}}
\tablewidth{0pt}
\tablehead{
\colhead{Obs. ID} & \colhead{Start date} & \colhead{Start time} &  \colhead{Turn-on} & \multicolumn2c{RGS1} & \multicolumn2c{RGS2} & \multicolumn3c{EPIC pn}  \\
\nocolhead{} & \nocolhead{} & \nocolhead{} & \nocolhead{} & \colhead{rate} & \colhead{exposure} & \colhead{rate} & \colhead{exposure} & \colhead{columns excluded} & \colhead{rate} & \colhead{exposure} \\
\nocolhead{} & \nocolhead{} & \colhead{MJD} & \colhead{MJD} & \colhead{s$^{-1}$} & \colhead{s} & \colhead{s$^{-1}$} & \colhead{s} & \colhead{} & \colhead{s$^{-1}$} & \colhead{s}
}
\startdata
0134120101 & 2001-01-26 & 51935.03629 & 51929.2 & 11.5 & 11300 & 2.48$^{1}$ & 11290$^{1}$ & 0 & 453 & 5660 \\
0153950301 & 2002-03-17 & 52350.01778 & 52348.8 & 3.49$^{2}$ & 7420$^{2}$ & 19.2 & 7260 & 1 & 488 & 2770 \\
0673510501 & 2011-07-31 & 55773.30710 & 55772.4 & 11.6 & 9530 & 12.6 & 9430 & 1 & 372 & 6800 \\
0673510601 & 2011-09-07 & 55811.32099 & 55807.2 & 16.8 & 32240 & 18.0 & 32000 & 2 & 379 & 19390 \\
0673510801 & 2012-02-28 & 55985.06984 & 55981.6 & 19.6 & 12800 & 21.1 & 12750 & 2 & 452 & 4940 \\
0673510901 & 2012-04-01 & 56018.83835 & 56015.8 & 11.7 & 13120 & 12.8 & 12960 & 1 & 387 & 9430 \\
0783770501 & 2016-08-17 & 57617.28046 & 57617.2 & 3.90 & 4950 & 4.20 & 4940 & 0 & 226 & 4930 \\
0783770601 & 2016-08-17 & 57617.82213 & 57617.2 & 9.56 & 5290 & 10.4 & 5290 & 1 & 342 & 4490 \\
0783770701 & 2016-08-18 & 57618.73301 & 57617.2 & 13.1 & 12330 & 14.3 & 12300 & 2 & 330 & 7040 \\
0830530101 & 2019-02-09 & 58523.53161 & 58516.6 & 10.9 & 19890 & 12.2 & 19860 & 1 & 419 & 13650 \\
0830530401 & 2019-03-14 & 58556.34063 & 58551.5 & 8.75 & 7330 & 9.70 & 7290 & 0 & 442 & 4020 \\
\enddata
\tablecomments{
$^{1}$values for RGS 1 2nd order spectra for observation 0134120101.
$^{2}$values for RGS 2 2nd order spectra for observation 0153950301.}
\end{deluxetable*}

\begin{deluxetable*}{ccccccccccc}
\tablecaption{Details of new \xmm\ observations used in this work. The high flux periods of the 3 full-orbit observations were split into 14 segments for detailed analysis of wind time evolution. All exposures listed are clean. The pn count rates shown are after pile-up correction. The measured Turn-on time used for all the August 2020 segments is MJD 59071.65.  \label{xcampdata}}
\tablewidth{0pt}
\tablehead{
\colhead{Obs. ID} & \colhead{Start date} & \colhead{Segment} & \colhead{Start time} & \multicolumn2c{RGS1} & \multicolumn2c{RGS2} & \multicolumn3c{EPIC pn}  \\
\nocolhead{} & \nocolhead{} & \nocolhead{} & \nocolhead{} & \colhead{rate} & \colhead{exposure} & \colhead{rate} & \colhead{exposure} & \colhead{columns excluded} & \colhead{rate} & \colhead{exposure} \\
\nocolhead{} & \nocolhead{} & \colhead{} & \colhead{MJD} & \colhead{s$^{-1}$} & \colhead{s} & \colhead{s$^{-1}$} & \colhead{s} & \colhead{} & \colhead{s$^{-1}$} & \colhead{s}
}
\startdata
0865440101 & 2020-08-10 & 1 & 59071.69035 & 4.8 & 4650 & 5.28 & 4610 & 0 & 230 & 2620 \\
 &  & 2 & 59072.33948 & 10.1 & 11070 & 11.1 & 11080 & 2 & 230 & 9590 \\
 &  & 3 & 59072.46795 & 10.9 & 11070 & 12.1 & 11090 & 2 & 247 & 9050 \\
 &  & 4 & 59072.59642 & 11.3 & 11060 & 12.5 & 11100 & 2 & 253 & 8820 \\
 &  & 5 & 59072.72489 & 11.6 & 11060 & 12.9 & 11090 & 2 & 259 & 8630 \\
 &  & 6 & 59072.85337 & 11.6 & 10960 & 12.8 & 10990 & 2 & 256 & 8570 \\
 &  & 7 & 59072.98068 & 11.9 & 8160 & 13.3 & 8000 & 2 & 264 & 6190 \\
0865440401 & 2020-08-12 & 8 & 59073.83832 & 13.3 & 20720 & 14.6 & 20520 & 2 & 284 & 20480 \\
 &  & 9 & 59074.07952 & 13.1 & 20850 & 14.4 & 20690 & 2 & 277 & 20600 \\
 &  & 10 & 59074.32073 & 13.2 & 20740 & 14.5 & 20780 & 2 & 281 & 20490 \\
 &  & 11 & 59074.56193 & 13.8 & 20840 & 15.3 & 20890 & 2 & 301 & 20590 \\
 &  & 12 & 59074.80313 & 13.9 & 20750 & 15.3 & 20590 & 2 & 307 & 20490 \\
0865440501 & 2020-08-14 & 13 & 59075.69364 & 12.9 & 42040 & 14.3 & 42124 & 1 & 380 & 27640 \\
 &  & 14 & 59076.18207 & 12.6 & 42090 & 14.0 & 41830 & 1 & 376 & 41580 \\
\enddata
\end{deluxetable*}

\vspace{-1.5cm}

\subsubsection{RGS}

The RGS data were reduced using the \textsc{rgsproc} task. Any background flaring periods exceeding the threshold count rate of 0.3 ct/s on CCD number 9 were excluded. There were no such periods during the August 2020 campaign. Due to the very high flux of Her X-1, the default RGS background regions were strongly contaminated by source counts. Therefore, we used the blank field background spectra for all \textit{High State} RGS datasets. The source flux is very high, up to 20 ct/s per detector, but there should not be significant pile-up in RGS data. Appendix A of \citet{Kosec+20} contains a calculation of the maximum Her X-1 count rate for each RGS CCD chip, which is found to be below the recommended pile-up thresholds for all CCDs. We primarily use 1st order grating spectra, and only analyze 2nd order spectra in a few cases where 1st order spectra are missing. Since RGS 2 was not working during archival observation 0134120101, we replaced its data by the 2nd order RGS 1 spectrum. Conversely, RGS 1 was not operational during observation 0153950301, and we replaced the spectrum with the 2nd order RGS 2 spectrum.

We use the 7 \AA\ (1.8 keV)  to 35 \AA\ (0.35 keV) wavelength range. The data are binned by a factor of 3 directly within the spectral fitting package \textsc{spex} \citep{Kaastra+96} to achieve an oversampling of the instrumental resolution by roughly a factor of 3. Wherever used, the 2nd order RGS spectra are binned by a factor of 6 to achieve similar sampling compared with 1st order spectra.

\subsubsection{EPIC pn}

The EPIC pn instrument was in Timing mode during all of the \textit{High State} observations. Nevertheless, many Her X-1 observations are still piled-up due to the extreme count rate, reaching as high as 800 ct/s in some observations. The data were reduced using regular routines with \textsc{epproc}. Only events of PATTERN$\leq$4 (single/double) were accepted for pn data. Due to the high source flux, regular background flaring period identification methods in EPIC pn are not valid. We identified any flaring periods by eye on a case-by-case basis, but periods where background flaring was important in comparison with source flux were very rare.

We mitigate pile-up by excluding 1 or 2 central pixel columns wherever necessary. The amount of excluded pixels is chosen so that the pile-up in the most affected spectral bins is at most 5\%. This strategy results in a spread of final pn count rates in different observations depending on the exact pointing of the satellite. In some cases the central two columns showed similar count statistics and both had to be removed (resulting in lower final count rates), in other cases one central column was exposed much more than the neighbouring ones and was the only one we had to remove (resulting in higher final count rates). The background regions were a few columns as far from the source on the chip as possible. Background was not important in any of the observations analyzed. 

The reduced spectra were binned to oversample the instrumental resolution by at most a factor of 3, and to at least 25 counts per bin using the \textsc{specgroup} routine. They were used in the energy range between 1.8 keV (7 \AA) and 10 keV. We found an issue with EPIC pn gain calibration, resulting in energy shifts of the order of 100 eV at $6-7$ keV. The issue, as well as our mitigation strategy is described in detail in Appendix \ref{app:pngain}.

\subsection{\chandra}

\chandra\ \citep{Weisskopf+02} observed Her X-1 13 times in the past, similarly to \xmm. However, only three observations occurred during the \textit{High State} and were of sufficient quality for this study. All these observations used the High Energy Transmission Gratings \citep[HETG,][]{Canizares+05}. One exposure happened in 2002, while the remaining two were part of the Her X-1 campaign of August 2020. The details of all \chandra\ observations are listed in Table \ref{chandata}. All observations were carried out in the Continuous Clocking mode to avoid pile-up.

\begin{deluxetable*}{ccccccc}
\tablecaption{Details of \chandra\ observations used in this work. The fourth and the fifth columns contain the MEG and HEG count rates obtained by stacking the positive and negative 1st orders of each instrument.\label{chandata}}
\tablewidth{0pt}
\tablehead{
\colhead{Obs. ID} & \colhead{Start date} & \colhead{Start time} &  \colhead{Turn-on} & \colhead{MEG rate} & \colhead{HEG rate} & \colhead{Clean exposure}  \\
\nocolhead{} & \nocolhead{} & \colhead{MJD} & \colhead{MJD} & \colhead{s$^{-1}$} & \colhead{s$^{-1}$} & \colhead{s} 
}
\startdata
2704 & 2002-07-05 & 52460.73963 & 52454.3 & 18.4 & 9.87 & 18650 \\
23356 & 2020-08-12 & 59073.22907 & 59071.65 & 11.6 & 7.97 & 17010 \\
23360 & 2020-08-14  & 59075.91068 & 59071.65 & 11.1 & 7.63 & 14830 \\
\enddata
\end{deluxetable*}

Observations 2704 and 23360 were downloaded from the TGCAT archive \citep[fully reduced,][]{Huenemoerder+11}. Her X-1 went into eclipse towards the end of observation 23356. Only the high flux part of the observation was extracted from 23356 by identifying the correct good time interval (GTI) times and applying the \textsc{chandra\_repro} routine \citep{Fruscione+06}. All observations were reduced using standard masks. We use data from both Medium Energy Grating (MEG) and High Energy Gratings (HEG), stacking the positive and negative first order spectra for each instrument using the \textsc{combine\_grating\_spectra} routine.

The data are analyzed in full spectral resolution without any binning. In the archival observation from 2002 (before the recent decline of \chandra's soft X-ray effective area), we use the MEG spectra between 0.6 keV (20 \AA) and 5 keV. During the new observations, we use MEG data in the 0.8 to 5.0 keV range. The HEG data are always used in the 1 to 9 keV range.

\section{Spectral Modelling} \label{sec:modelling}

All reduced spectra were converted from \textsc{ogip} format into \textsc{spex} format using the \textsc{trafo} routine. The fitting was done in the \textsc{spex} fitting package \citep{Kaastra+96} using Cash statistics \citep{Cash+79}. The \textsc{spex} usage was motivated by the \textsc{pion} model \citep{Miller+15, Mehdipour+16}, which we used to describe the ionized absorption from the Her X-1 accretion disk wind. \textsc{pion} is a photoionization spectral model which can describe both photoionized emission and absorption. The ionization balance is calculated on the go within \textsc{spex} using the currently loaded continuum model as the spectral energy distribution. 

We use cross-calibration constants to account for any minor calibration differences between the \xmm\ RGS1, RGS 2 and EPIC pn instruments. RGS 1 and RGS 2 spectra were not stacked, and are fitted simultaneously with the EPIC pn spectra. We find that RGS 1 and 2 agree within 2\%, and EPIC pn and RGS agree within 7.5\% in most cases. Similarly, we use cross-calibration constants when simultaneously fitting \chandra\ MEG and HEG spectra (6\% maximum difference).

In the subsections below we describe the various spectral components and their properties before proceeding to the results. Thanks to the very high flux of Her X-1 and the long exposures, the data quality is extremely high with millions of counts in each observation or observation segment. However, the spectrum of Her X-1 is also very complex and requires many spectral components. As our main goal for the program is to study the absorption lines from the disk wind, the broadband continuum must be modelled accurately to prevent the absorber models from fitting residuals due to incorrect continuum modelling instead of real disk wind features.

We adopt phenomenological models for the various emission components, and apply a physical model for the ionized absorption from the accretion disk wind (\textsc{pion}). A simple continuum modelling is required to reduce the computational cost of the fitting analysis, particularly due to the usage of the \textsc{pion} model.

As noted previously, we found an issue with the gain calibration of the EPIC pn spectra. The effect is that pn photon energies appear to be too high, by as much as a factor of 1.015-1.02. The issue is described in greater detail in Appendix \ref{app:pngain}. Our solution is to systematically blueshift all spectral models fitting the EPIC pn spectra with a \textsc{reds} component within \textsc{spex}. The RGS spectra have the same \textsc{reds} component applied but its blueshift is set to 0. The actual EPIC pn gain shift value is unknown but we use several fixed parameters to anchor our spectral model between the RGS and the EPIC pn data. These parameter include some Fe K region emission lines (described in more detail below) as well as the wavelengths of the ionized wind absorption lines. The absorption lines are modelled with a single \textsc{pion} photoionization grid using only one blueshift parameter, thus coupling their line positions between the RGS and EPIC pn energy bands.

Secondly, we found a broad residual around 23\AA\ in the RGS band which affected our broadband continuum fit. The residual could be due to instrumental calibration or imperfect modelling of interstellar hot gas and dust. It is described in more detail in Appendix \ref{app:ignrgs}. To simplify the spectral modelling, in this paper we thus ignore the affected wavelength range between 22.25 and 24 \AA\ ($0.52-0.56$ keV).

\subsection{Broadband spectrum of Hercules X-1}

The hard X-ray ($>2$ keV) spectrum of Her X-1 is dominated by the pulsating emission from the accretion column. The emission can be described as a powerlaw with a cutoff of $\sim20$ keV (beyond the reach of \xmm\ and \chandra\ bandpass). Here we model the accretion column emission with a \textsc{comt} comptonization model in \textsc{spex} \citep{Titarchuk+94}.

The soft X-ray ($<1$ keV) spectra of the \textit{High State} are instead dominated by soft excess resembling a black body with a temperature of 0.1 keV \citep{Hickox+04}. This component pulsates at the neutron star rotation period as well \citep{Ramsay+02} and thus does not correspond to accretion disk thermal emission. The blackbody is likely accretion column beam radiation reprocessed off the warped disk \citep{brumback+21}. We attempted to model the soft X-rays of the high quality new \xmm\ observations with a single blackbody of $\sim0.1$ keV temperature but could not recover a reasonable broadband continuum fit. This is due to the overall spectral shape being broader than a single blackbody \citep[previously also found by][]{Kosec+20}.  We note that our soft X-ray coverage is resolved in high spectral resolution with the RGS detectors and thus the broad continuum shape is not due to resolution issues. To achieve a better fit with our phenomenological model, we introduce a broader blackbody shape, the \textsc{mbb} model \citep[blackbody modified by coherent Compton scattering,][]{Kaastra+89}, as well as a second, regular blackbody. This results in significant fit quality improvements (\delcstat$>100$). 

Physically, the second blackbody could correspond to accretion disk thermal emission while the hotter blackbody is reprocessed emission of the primary accretion column beam radiation. Alternatively, it may be that the reprocessed column radiation can not actually be perfectly described with a single blackbody spectrum. The temperature of the second blackbody is lower than that of the first, around 0.05 keV or even less. It is thus difficult to constrain it with our RGS data which end at 35\AA\ (0.35 keV). For this reason, we fix the second blackbody to 0.05 keV (the best-fitting temperature wherever it can be constrained) to avoid it running away to too low temperatures with unphysically large emitting areas and luminosities (below our energy coverage).

The soft X-ray spectra of Her X-1 around 1 keV also show another spectral component named the `broad 1 keV feature' \citep{Fuerst+13}. The component is further shown at high spectral resolution in Fig. 2 of \citet{Kosec+20}, as well as in our new high quality spectra (Fig. \ref{401_RGS}). Even at RGS resolution, the feature is reasonably well fitted with a single broad Gaussian with a width of $0.35-0.4$ keV and it does not appear to be composed of a smaller number of narrow features. The component is plausibly a blend of a forest of Fe L and Ne IX-X lines \citep[possibly similar to the broad Fe L emission lines seen in some AGN,][]{Fabian+09} but physical models need to be applied to understand its nature properly. Here, as a preliminary study, we track its evolution phenomenologically with a simple Gaussian component.

\subsection{The complex Fe K band}

The \textit{Low State} Her X-1 spectrum contains a strong and narrow ($<0.1$ keV) emission line very close to the neutral Fe energy of 6.4 keV \citep{Jimenez+05}, possibly originating on the irradiated face of the secondary HZ Her \citep{Zane+04}. In the \textit{High State} instead the Fe emission is much broader ($0.3-0.5$ keV) and located at higher energies, closer to the Fe XXV transition at 6.67 keV \citep{Ramsay+02, Zane+04}. In our previous study \citep{Kosec+20}, we modelled this region with a single Gaussian line, finding rest-frame energies of 6.5-6.6 keV and widths as large as 1 keV (full width half maximum, FWHM). Other studies suggest that the Fe K band is more complex, finding evidence for two or even three Gaussian components of various widths including a very broad component (FWHM$\sim2$ keV) \citep{Fuerst+13, Asami+14}. On top of the emission features are imprinted the narrow absorption lines from the disk wind \citep{Kosec+20}, leading to a complex spectral region.

\begin{figure}
\includegraphics[width=\columnwidth]{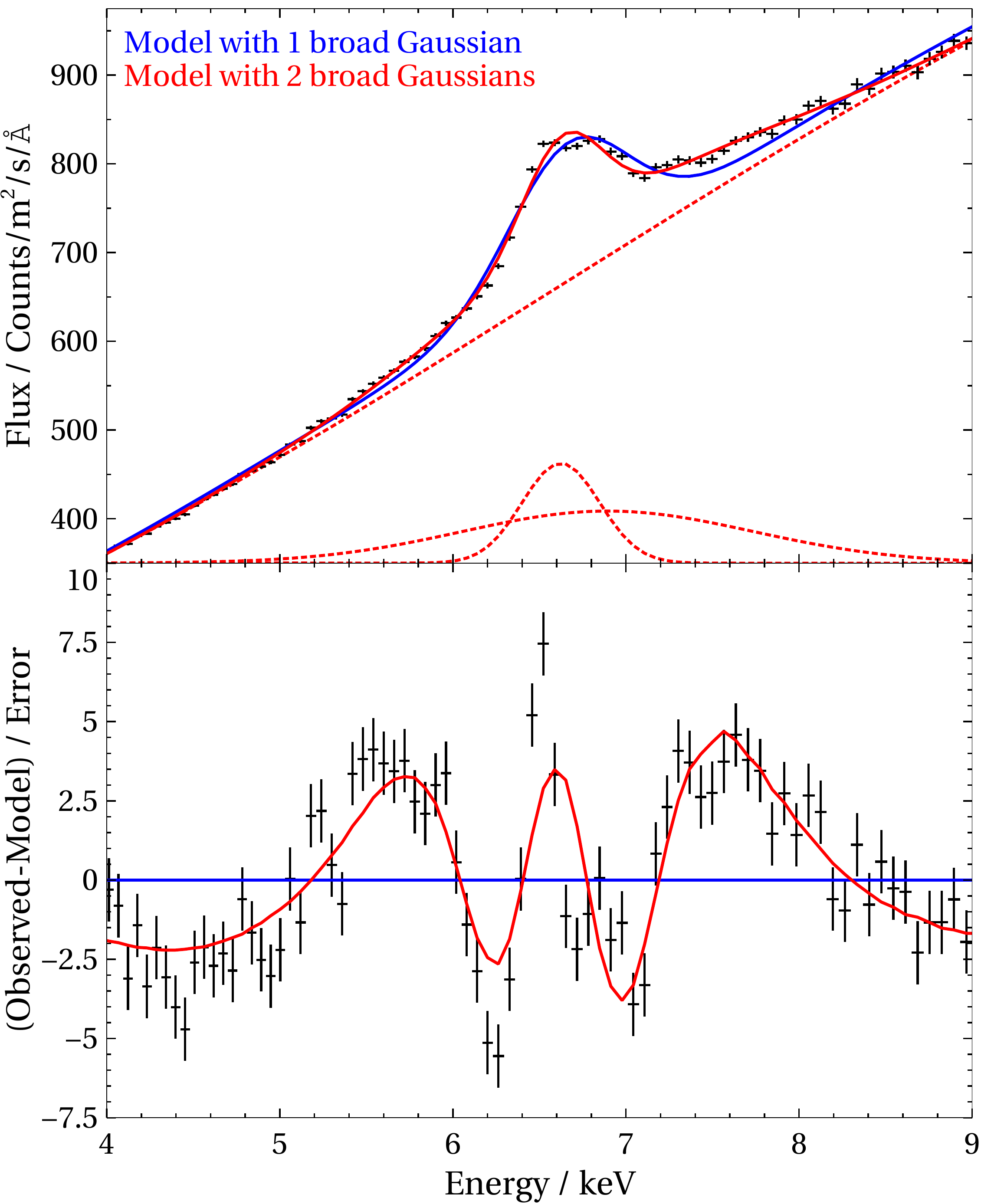}
\caption{EPIC pn spectrum of the \textit{High State} from observation 0865440501 (80 ks clean exposure), focusing on the Fe K region. The data are shown in the top plot in black (errorbars are so small they are difficult to see in the spectrum). The blue curve shows the best-fitting model containing one broad Gaussian, while the red curve corresponds to the model containing two broad Gaussian lines. The red dashed lines in the top subplot are the individual spectral components in the more complex, two Gaussian model. The Gaussian components are shifted by a constant amount for better visual clarity. The bottom subplot shows the residuals to the blue model in black, as well as how the model with two Gaussians fits the data (in red).\label{FeKxmmres}}
\end{figure}

We initially fitted the high quality \xmm\ spectra from the new 2020 campaign with a single Gaussian for the Fe feature, but discovered strong residuals in the Fe K band indicating that the single Gaussian is indeed insufficient. This is shown in Fig. \ref{FeKxmmres}. Instead of a single Gaussian with an energy of 6.6 keV and a FWHM of 0.8 keV, a much better fit (by \delcstat$\sim700$ for unsegmented observation 0865440501) is obtained by adding two Gaussians, a medium-width one at 6.6 keV with a FWHM of 0.4-0.5 keV and a second, broad one at 6.7 keV with a FWHM width of 1.9 keV (best-fitting energies are given without any gain correction here). The broad Gaussian indicates extreme plasma velocities of 0.1-0.15c. However, the line properties are similar to what \citet{Fuerst+13} and \citet{Asami+14} found using \nustar\ and \suzaku\ data, respectively (thus arguing against the EPIC pn residual being of instrumental origin).

\begin{figure}
\includegraphics[width=\columnwidth]{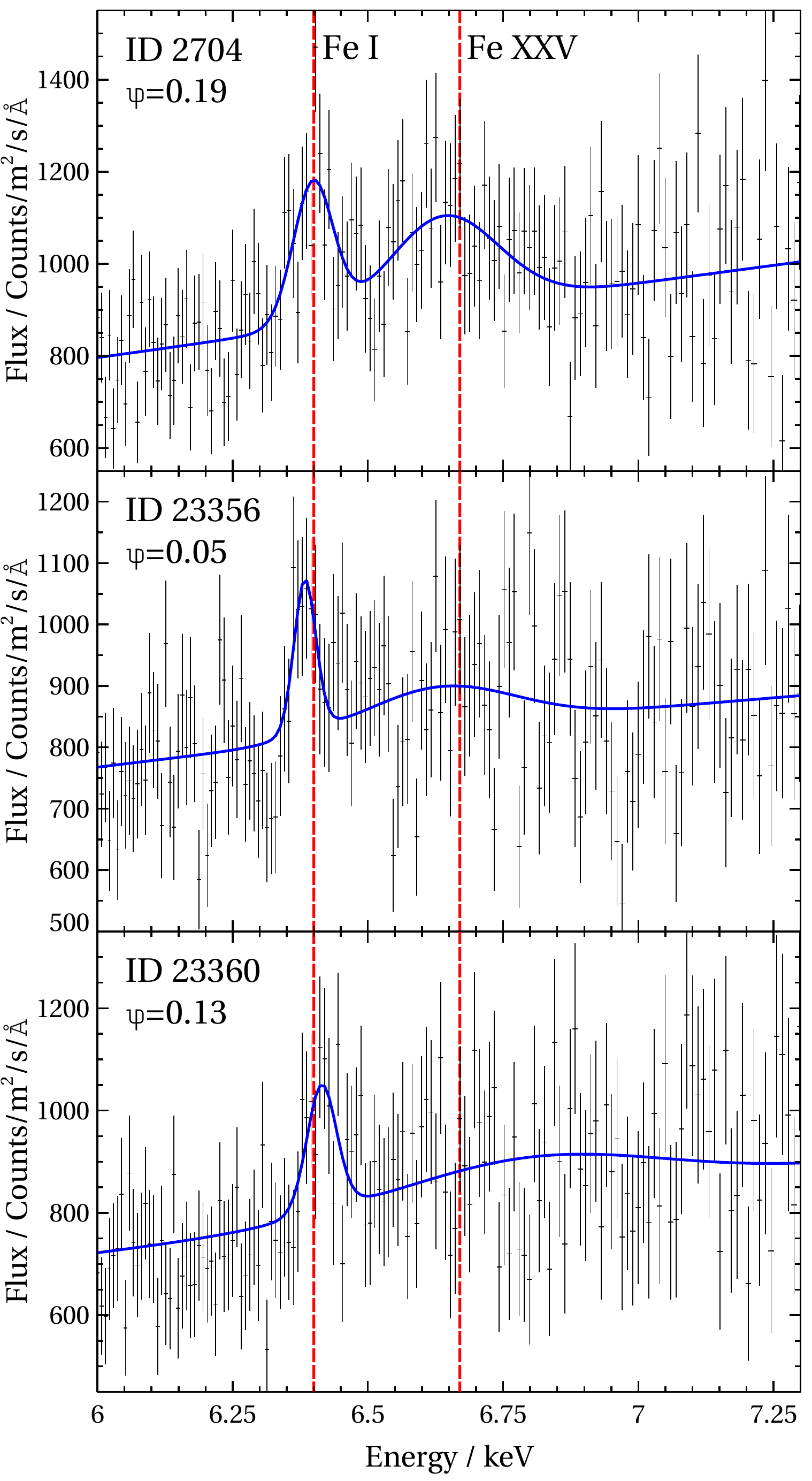}
\caption{Spectra from the three \chandra\ observations of Her X-1 \textit{High State}, focusing on the Fe K band. The data are fitted with a simple spectral model consisting of a powerlaw plus two Gaussians, shown in blue. The red dashed lines show the positions of Fe I (6.4 keV) and Fe XXV (6.67 keV) transitions. The super-orbital phase $\phi$ of each \chandra\ observation is given below its ID.  \label{FeKchandra}}
\end{figure}

Fig. \ref{FeKxmmres} hints that the Fe K band is likely even more complex. A sharp residual at around $6.4-6.5$ keV is seen in the more complex 2-Gaussian spectral model, indicating the presence of a narrow (less than $0.1-0.2$ keV width) emission line close in energy to neutral Fe transition. However, it is difficult to constrain this narrowest emission line with the limited energy resolution of the EPIC instrument. We therefore turn to the HETG observations of Her X-1 \textit{High State}. HETG spectra have poorer statistics due to a lower effective area than EPIC pn but offer significantly better spectral resolution. The spectra of the three \chandra\ observations, focusing on the Fe K band, are shown in Fig. \ref{FeKchandra}.

The HETG resolution is sufficient to resolve the emission lines. A narrow line near 6.4 keV is clearly resolved, which does not appear to vary significantly in position or width. Its FWHM width is around 50 eV. This is consistent with the Fe I transition energy and thus we denote this feature `Fe I' in the text below. However, we note that this component could also contain line emission from other low ionization Fe ions (up to at least Fe XIII), the transition energies of which are difficult to distinguish from 6.4 keV using our data. A second, medium-width Gaussian is also present, and is consistent with the Fe XXV transition at 6.67 keV. The width of this feature appears to be highly variable. Any broad emission component with FWHM$\sim1.5-2$ keV is not statistically significant in any of the Chandra observations (regardless of spectral binning) due to lower count statistics in the HETG spectra.

Based on the findings above, we build our spectral model for the Fe K band of Her X-1, which contains three emission lines. A narrow line at 6.4 keV, with a FWHM width of about 50 eV, a medium-width line at around 6.67 keV, with a FWHM width of $\sim$0.5 keV (but highly variable between the observations), and a very broad line around 6.5 keV with a FWHM width of 1.5-2.0 keV. In the \chandra\ analysis, the last line is not present in the model and all the line positions and widths are left free to vary. In the \xmm\ analysis, we fix the width of the narrowest Gaussian to 50 eV ($\sim1000$ km/s velocity width) as it is too narrow for the EPIC pn spectral resolution. Unfortunately, we cannot fit for the two narrower line energies using \xmm\ observations due to issues with EPIC pn gain (Appendix \ref{app:pngain}). Instead, we fix the energies to 6.4 and 6.67 keV and use the features as one of the anchors for the EPIC pn spectrum. Finally, the broadest feature, which does not have an obvious identification, has both the energy and width left free to vary, although we require the energy to be at least 6.4 keV.

\subsection{The soft X-ray lines}

\begin{figure*}
\includegraphics[width=\textwidth]{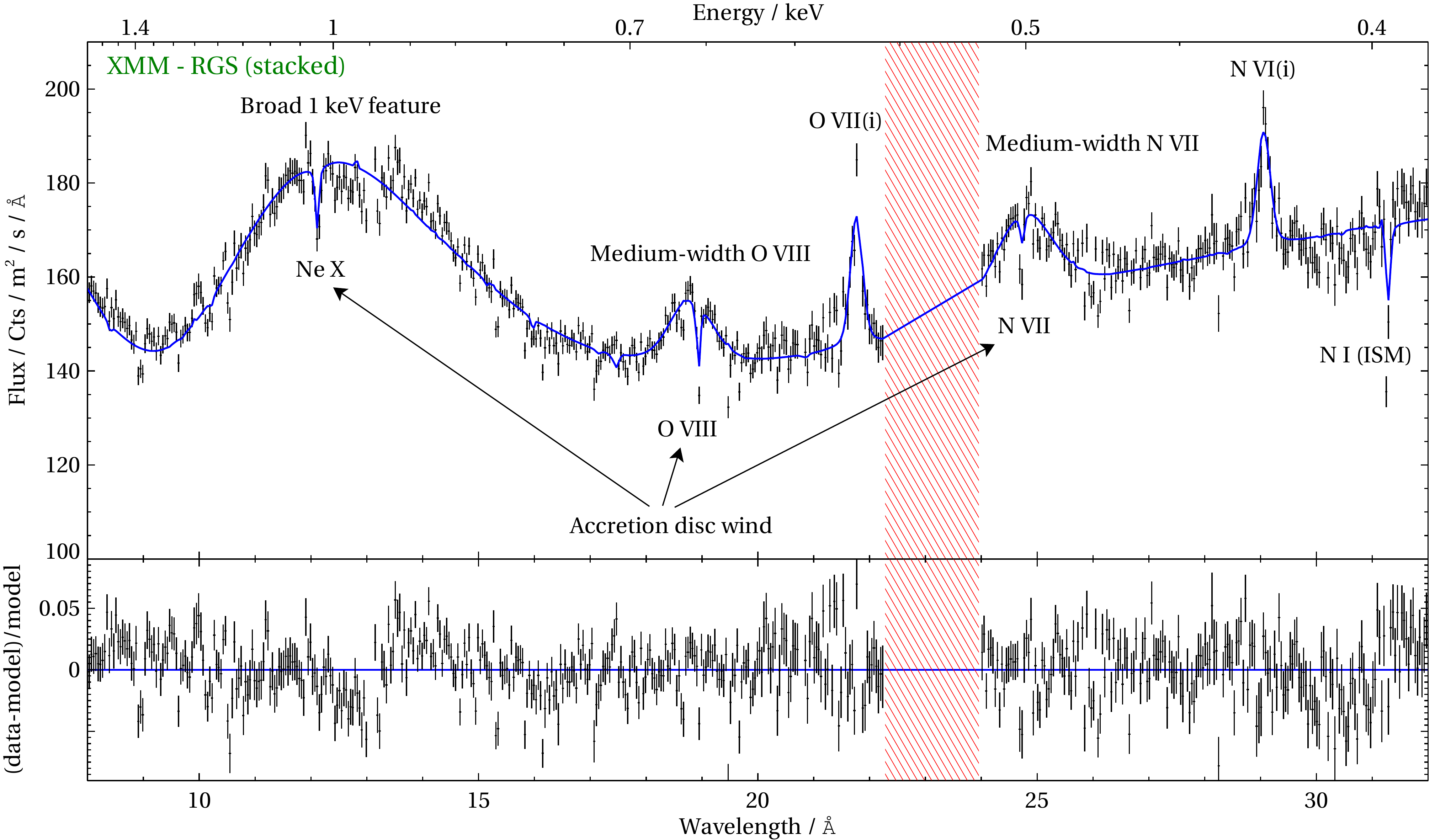}
\caption{RGS spectrum (RGS 1 and 2 stacked for visual purposes) from the full 100ks long observation 0865440401 (top panel), fitted with the full spectral model. Notable spectral components and features are described with labels. The red shaded region was ignored in this analysis (Appendix \ref{app:ignrgs}). The bottom subpanel shows the ratio residuals to the best-fitting model. \label{401_RGS}}
\end{figure*}

The soft X-ray ($<$1 keV) \textit{High State} spectrum contains several strong emission line features. An example high quality RGS spectrum (from observation 0865440401) is shown in Fig. \ref{401_RGS}. These spectral features have also been previously discussed in our first paper \citep{Kosec+20}.

Medium-width ($5000-10000$ km/s) O VIII and N VII emission lines are apparent in the RGS spectra. These were previously modelled with a collisional ionization equilibrium model (\textsc{cie}). However, with the new high quality data from 2020 we noticed that the model consistently over-estimated the line widths (reaching as high as $15000-20000$ km/s velocity widths). Here we simplify the model for these features to two Gaussians, with the velocity width tied between the two components. The velocity widths appear to range between 5000 and 10000 km/s using this new model. We also do not fix the wavelengths of the two features to the rest-frame transitions of O VIII and N VII as we notice clear deviations from those positions in some \xmm\ observations.

It is crucial to describe these medium-width emission features correctly as the narrow O VIII and N VII absorption lines from the disk wind are located on top of these broadened features. Over-predicting the line widths would then reduce the recovered narrow absorption line depths.

The soft energy band also contains narrower emission lines with widths of $\sim$1000 km/s. They most likely originate in the so-called accretion disk corona (ADC) \footnote{We note that the ADC is a different type of region than a typical accreting black hole corona, which is located very close to the compact object.} above the disk \citep[see Fig. 9 of][]{Jimenez+02, Kuster+05}. The most prominent are the intercombination lines of O VII and N VI, which we model with Gaussian lines. There is also evidence for the intercombination line of Ne IX at 13.55 \AA, but the line is not very prominent in the individual observational segments. Therefore we do not model it here in order to decrease the computational cost of the fitting procedure. We free all the O VII and N VI Gaussian parameters, but require their widths to be at most $FWHM=0.5$ \AA\ (velocity widths at most of roughly 3000 km/s). This is to prevent the widths running away to too large values and the Gaussians acting as broadband continuum components instead. 

From the residual plot in Fig. \ref{401_RGS}, it can be seen that many more, weaker features are still present in the spectrum. These features are currently not modelled as the present initial study focuses only on the strongest components of the soft X-ray spectrum. The weaker residuals will be described and studied in depth in a future publication.

\subsection{The final spectral model}

To summarize the subsections above, the final spectral continuum model used for Her X-1 \textit{High State} analysis is as follows. The broadband components are described with a \textsc{comt} Comptonization component (primary accretion column radiation), \textsc{mbb} and \textsc{bb} blackbody components (accretion column reprocessed radiation) and a broad Gaussian residual near 1 keV (possibly broad Fe L and Ne IX/X emission). The Fe K band is modelled using three Gaussian components of various widths (Fe I, Fe XXV and very broad Fe K). The soft X-ray band contains two further medium-width Gaussians (O VIII and N VII) and two narrow Gaussians (O VII and N VI intercombination lines). For each of the spectral components as well as for the full model, we calculate the $0.01-80$ keV observed luminosity using \textsc{spex}, assuming a 6.1 kpc distance of Her X-1 \citep{Leahy+14}.

On top of this continuum we apply the photoionized absorber model \textsc{pion} which describes the accretion disk wind features. \textsc{pion} calculates the ionization balance from the SED of the spectral components and self-consistently determines the optical depths of the absorption lines. We fit for the column density, ionization parameter, velocity width as well as the blueshift of these absorption lines, while fitting for the non-Solar abundances of the Her X-1 system. More details of the \textsc{pion} modelling as well as the results are given in Paper II.

All of these components are obscured by Galactic neutral absorption, modelled using a \textsc{hot} component. The neutral column density towards Her X-1 is very low, around $1.5 \times 10^{20}$ \pcm\ \citep{HI4PI+16}, however many of our fits prefer even lower column values. This is understandable since Her X-1 is located within our Galaxy and the given neutral column density is the integral of all neutral gas along the line of sight towards (and beyond) the Her X-1 position. We set a lower limit on the column density of $1.0 \times 10^{20}$ \pcm.

We correct the EPIC pn gain issue by applying a \textsc{reds} component which blueshifts the EPIC pn spectral model by a multiplication factor. Further details of this gain correction are shown in Appendix \ref{app:pngain}.


The final spectral model for \xmm\ data is therefore, in symbolic form: \textsc{reds}$\times$\textsc{hot}$\times$\textsc{pion}$\times$[\textsc{comt} + \textsc{mbb + bb} + \textsc{gaus}(Fe L) + 3$\times$\textsc{gaus}(Fe K) + 4$\times$\textsc{gaus}(O VIII, N VII, O VII and NVI)]. All components and parameters of this spectral fit are summarized in Table \ref{paramtable} (Appendix \ref{app:fitpar}). The model for \chandra\ data with its more limited statistics and no gain issues is significantly simpler, especially in the soft X-ray band: \textsc{hot}$\times$\textsc{pion}$\times$[\textsc{comt} + \textsc{mbb} + \textsc{gaus}(Fe L) + 2$\times$\textsc{gaus}(Fe K) + \textsc{gaus}(O VIII)].

\section{Results}\label{sec:results}

This section contains the results of time-resolved analysis of all the new and archival \xmm\ and \chandra\ Her X-1 \textit{High State} observations. Many archival \xmm\ observations occurred years before the August 2020 observational campaign (which captured a part of a single disk precession cycle). Possible long-term variations in the accretion flow of Her X-1 could introduce systematic differences between the archival and the 2020 results. For this reason, we show the archival and the new 2020 \xmm\ observation measurements apart in all relevant figures below (by using different colours for these measurements). In general, we find that the archival results are consistent with the 2020 results, indicating a very stable accretion flow in Her X-1, with limited long-term systematic parameter variations or trends over the past 20 years. This is in agreement with the observed stable super-orbital 35 day cycle in the past 20-year period, with the exception for the anomalous low state that occurred in 2003 \citep{Leahy+20}.


We focus on the properties and variability of the various emission line components, modelled as described in the previous section. Our primary goal is to show how these plethora of lines are related to each other, and determine their likely origin and location within the accreting system. We want to test if the different line properties correlate more with the observed Her X-1 luminosity (i.e. they are varying in line with the X-ray continuum flux, which originates in the inner accretion flow), the super-orbital phase (i.e. their changes are largely driven by geometric effects due to warped disk precession) or the orbital phase (i.e. they are somehow related to the secondary HZ Her). 


We make an important note about the measured X-ray luminosities. The X-ray continuum or line flux variations in Her X-1 do not correspond to intrinsic accretion luminosity variations (e.g. due to time variable mass accretion rate onto the neutron star). Even during the \textit{High State}, most of the flux variability is likely due to obscuration (column density or covering fraction) changes, either by the warped accretion disc, or by the ionized accretion disk atmosphere \citep{Scott+00}. Therefore, all the measured X-ray luminosities (of the continuum or the emission lines) are strictly the \textit{observed}, apparent luminosities obtained by converting the flux values using the distance of Her X-1 of 6.1 kpc, and should not be taken at face value as intrinsic luminosities. We list these apparent luminosity values in the units of erg/s rather than using the raw X-ray fluxes to allow a more straightforward comparison to other X-ray binaries.

\subsection{The narrow emission lines}



We focus on the three strongest narrow lines: Fe I, O VII and N VI with similar, low line widths.

\begin{figure}
\includegraphics[width=\columnwidth]{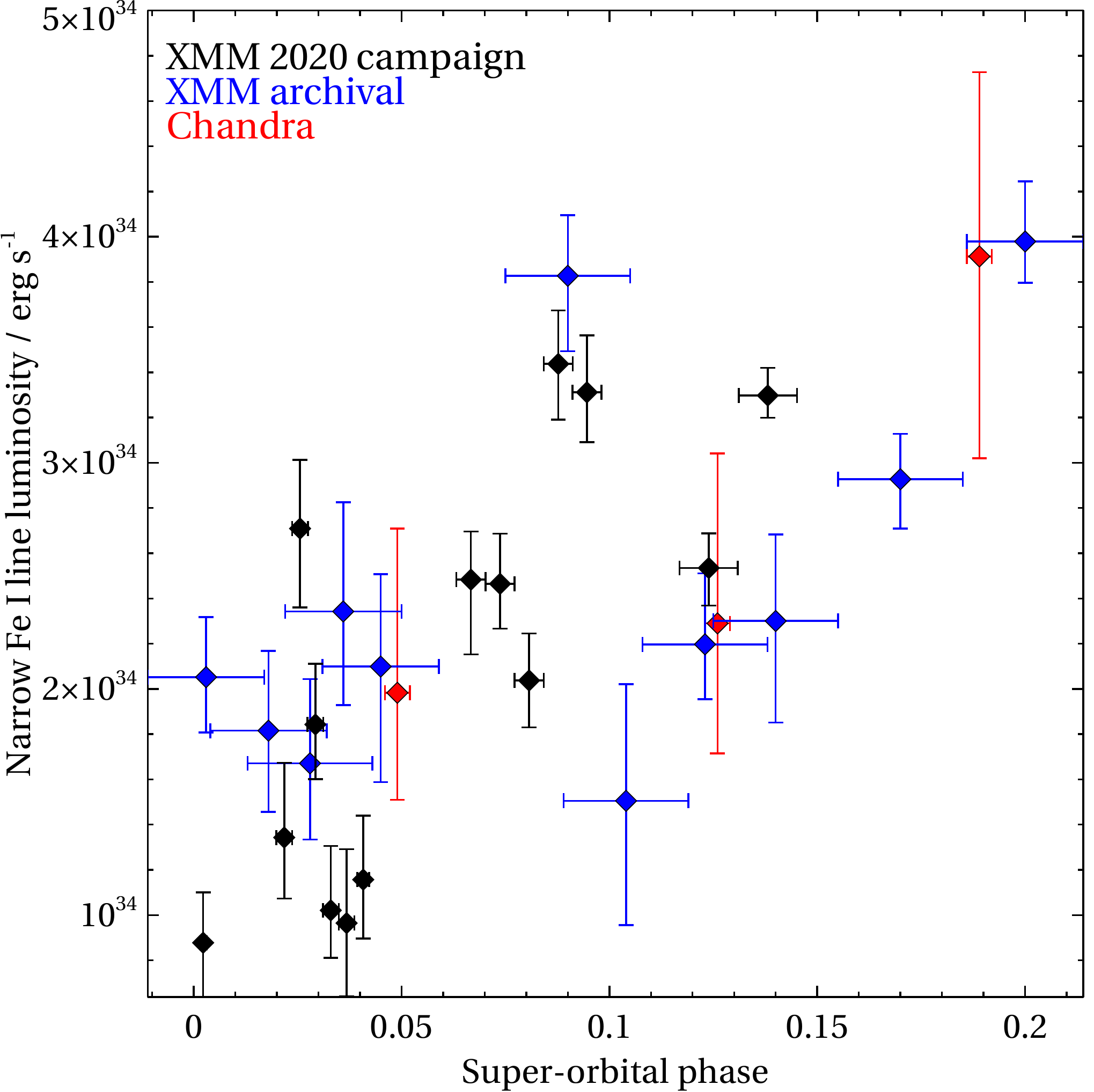}
\caption{The observed luminosity of the narrow Fe I line versus the super-orbital phase. The observations with \xmm\ during the 2020 campaign, or previously, as well as \chandra\ observations are shown in distinct colours. \label{FeI_lum}}
\end{figure}

The luminosity of the Fe I line varies in the range $(1-4)\times10^{34}$ erg/s (about 1/1000 of the observed Her X-1 luminosity). We find that it significantly correlates with the super-orbital phase of Her X-1 (Fig. \ref{FeI_lum}, Pearson correlation coefficient 0.70, p-value $3.5\times10^{-5}$), much more than with the observed Her X-1 luminosity (Pearson coefficient 0.17). We do not observe any significant correlations with the orbital phase during the \textit{High State} (Pearson coefficient -0.32). The correlation with super-orbital phase appears especially strong in the new August 2020 observations (Pearson coefficient 0.76), taken over the course of a single precession cycle (thus avoiding confusion with any possible long-term variations in the Her X-1 system).

The soft X-ray spectra of Her X-1 \textit{High State} contain strong intercombination lines of the O VII and N VI He-like triplets. We do not find clear correlations between the observed O VII(i) and N VI(i) line luminosities and Her X-1 orbital or super-orbital phase. There is a positive trend of O VII(i) and N VI(i) luminosity with the observed Her X-1 luminosity but with many outliers (Pearson correlation coefficients 0.56 and 0.55, respectively).

\begin{figure*}
\includegraphics[width=\textwidth]{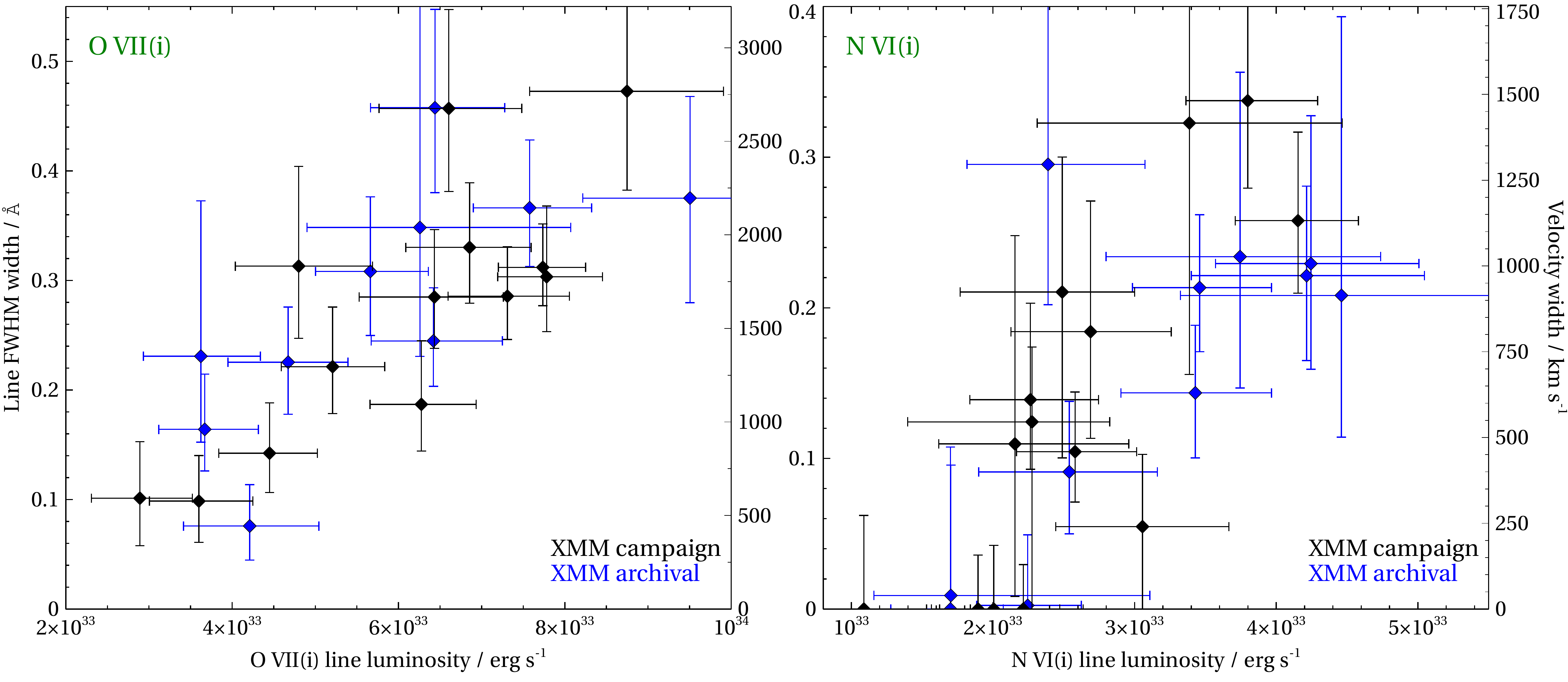}
\caption{Left panel: The best-fitting FWHM line width of the O VII(i) line versus its observed luminosity. Right panel: FWHM line width of the N VI(i) line versus its observed luminosity. The left Y-axes on each panel show the line widths in \AA, while the right Y-axes show the velocity widths in km/s. \label{OVII_NVI}}
\end{figure*}

While these features are very narrow during the \textit{Low State} \citep[$<$300 km/s,][]{Jimenez+05}, we find that they are significantly broader during the \textit{High State}, as well as about $3-5$ times brighter. The measured line widths are found to be highly correlated with the observed line luminosities, and are shown in Fig. \ref{OVII_NVI}. The Pearson correlation is 0.76 (p-value $2.8\times10^{-5}$) for O VII(i) and 0.74 (p-value $2.4\times10^{-5}$) for N VI(i). The N VI(i) width is less than 0.1 \AA\ ($<400$ km/s velocity width) at the lowest luminosities, reaching up to 0.3 \AA\ (1300 km/s) at maximum. The O VII(i) feature is broader, with a width of at least 0.1 \AA\ (600 km/s) at minimum fluxes, and reaching widths of 0.5 \AA\ (3000 km/s) at maximum fluxes.

We find that the O VII(i) best-fitting wavelength is weakly anticorrelated with its observed luminosity (Pearson correlation -0.50, p-value $1.5\times10^{-2}$), beginning at its rest-frame wavelength of 21.8 \AA\ for low O VII(i) luminosities but reaching as low as 21.7 \AA. This likely indicates some contamination with the O VII resonance line at 21.6 \AA\ (0.2 \AA\ away from the O VII(i) line). The O VII(r) line, clearly weak at low O VII(i) luminosities, could increase in importance, pulling down the wavelength of our single Gaussian - this is not a spectral resolution issue but a modelling problem. Therefore, the O VII(i) width could be over-estimated at the largest line luminosities, where it appears the broadest (FWHM$>0.3$ \AA). Nevertheless, in observations with lower O VII(i) widths where no contamination occurs (best-fitting line position at 21.8 \AA\ with FWHM widths of $0.1-0.2$ \AA), the line broadening of $600-1200$ km/s is still much larger than observed in the \textit{Low} and \textit{Short High} states by \citet{Jimenez+02}.

The N VI(i) line does not suffer from the same issue as O VII(i), thanks to both its smaller broadening as well as due to the higher wavelength separation from the N VI(r) line ($\sim$0.3 \AA). The Pearson correlation coefficient of N VI(i) observed luminosity and wavelength is -0.26 (p-value 0.21), indicating low correlation between the two parameters. We still observe a similar positive trend of the line broadening with the line luminosity, as well as some variations around the best-fitting line position. The wavelength variations are at most 0.05 \AA, smaller than the changes in the contaminated O VII(i) line wavelength. We note that the expected variations due to the neutron star orbital motions should be significantly lower at about 0.02 \AA\ around the line rest-frame.

\subsection{The medium-width emission lines}

\begin{figure}
\includegraphics[width=\columnwidth]{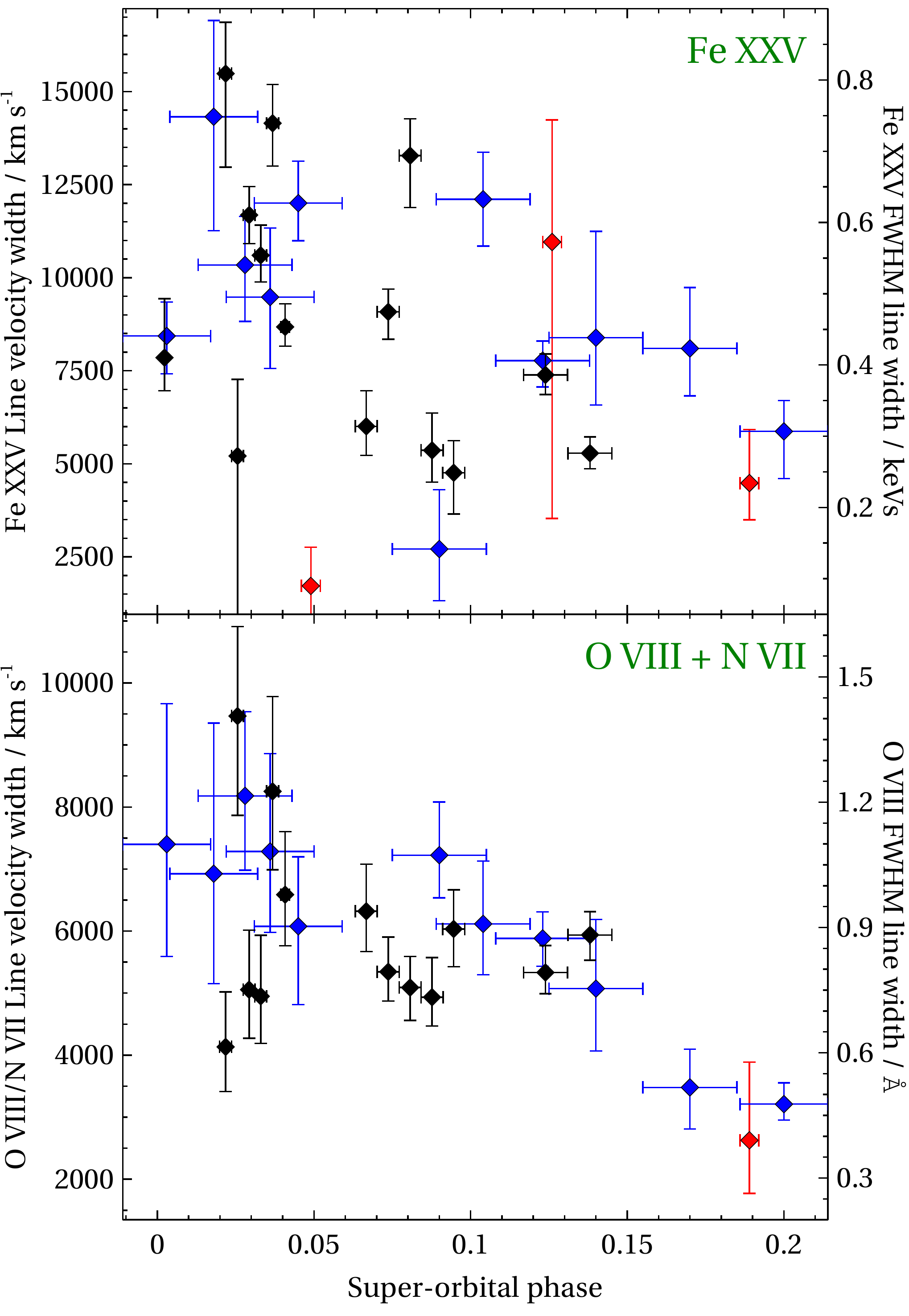}
\caption{The 1D velocity (left Y-axes) and FWHM widths (right Y-axes) of the Fe XXV (top panel) and O VIII/N VII (bottom panel) lines versus the super-orbital phase. The right Y-axis on the bottom panel only contains the FWHM width of the O VIII line. The observations with \xmm\ during the 2020 campaign are shown in black, archival \xmm\ observations are in blue, and \chandra\ observations are in red colour. \label{FeXXV_OVIII_width}}
\end{figure}

We resolve the Fe K band of Her X-1 into multiple line components using \chandra\ HETG and \xmm\ EPIC spectra, and constrain their widths. \chandra\ observations hint at strong variability particularly in the broadening of the medium-width Fe XXV feature. The FWHM width of the line is indeed strongly variable, between $\sim$0.2 keV (4000 km/s) and as high as 0.8 keV (15000 km/s). There is a negative trend of the width with the super-orbital phase, shown in Fig. \ref{FeXXV_OVIII_width} (top subplot), but with many outliers (Pearson coefficient -0.38, p-value $4.7\times10^{-2}$). We find a possible weak correlation between the Fe XXV luminosity and the observed Her X-1 luminosity (Pearson coefficient 0.43, p-value $2.3\times10^{-2}$).

The medium-width O VIII and N VII lines were detected and briefly described by \citet{Kosec+20} but that study focused on the disk wind absorption. We note that the velocity widths of the two features are tied in the spectral fit, but their wavelengths, and their normalizations are decoupled.

We find that the observed luminosities of the two lines are highly correlated (Pearson coefficient 0.78, p-value $4.9\times10^{-6}$), indicating that they very likely have a common origin. There are no tight correlations between the line luminosities and Her X-1 super-orbital phase or orbital phase. On the other hand, we do find positive trends with the observed Her X-1 luminosity for both O VIII (Pearson coefficient 0.76, p-value $1.0\times10^{-5}$) and N VII (Pearson coefficient 0.73, p-value $4.0\times10^{-5}$). There is also a negative trend of the width of the lines with the super-orbital phase (Fig. \ref{FeXXV_OVIII_width}, bottom subplot). At the beginning of the precession cycle, the widths (of O VIII) are as large as 1.4 \AA\ (velocity width of 10000 km/s), while towards the end of the \textit{High State} they are much lower, at around 0.5 \AA\ (velocity widths of 3500 km/s). We therefore observe similar trends in line width evolution as with the Fe XXV line.

Finally, we also detect a curious evolution of the best-fitting wavelengths of the two lines with the super-orbital phase. This cannot be explained by a gain shift issue since these lines are in the RGS energy band, and RGS does not suffer from the same gain problem as pn. The O VIII line is initially redshifted from the rest-frame transition by about 0.2 \AA\ ($\sim$3000 km/s), then later systematically blueshifted (by up to 1500 km/s), followed by a return back to the rest-frame wavelength, and tending towards redshift at the end of the \textit{High State} again. The N VII line (which is more poorly constrained than O VIII) is systematically redshifted throughout the \textit{High State} by as much as 0.5 \AA\ (6000 km/s), but appears to be trending towards the rest-frame wavelength around the end of the \textit{High State}. We note that the shifts are always lower than the line FWHM widths, and their identification is secure as they are by far the strongest features of these widths in the Her X-1 RGS spectra.

\begin{figure}
\includegraphics[width=\columnwidth]{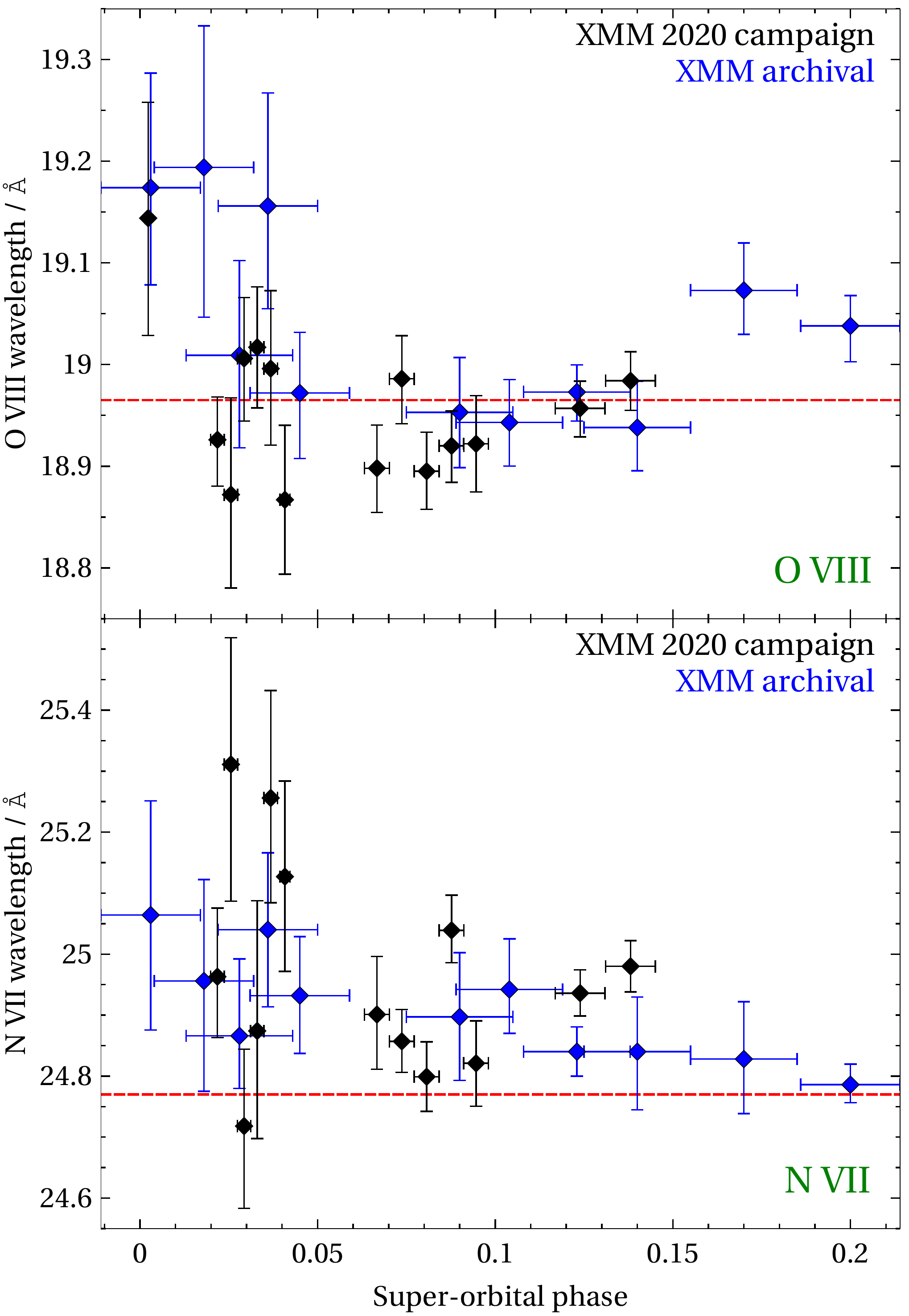}
\caption{The best-fitting wavelength of the O VIII line (top subplot) and the N VII line (bottom subplot) versus the super-orbital phase. The observations with \xmm\ during the 2020 campaign, and previously are shown in distinct colours. The red dashed horizontal lines indicate the rest-frame wavelengths of the two transitions. \label{OVIII_NVII_wave}}
\end{figure}

\subsection{The broad Fe L and Fe K lines}



\begin{figure*}
\includegraphics[width=\textwidth]{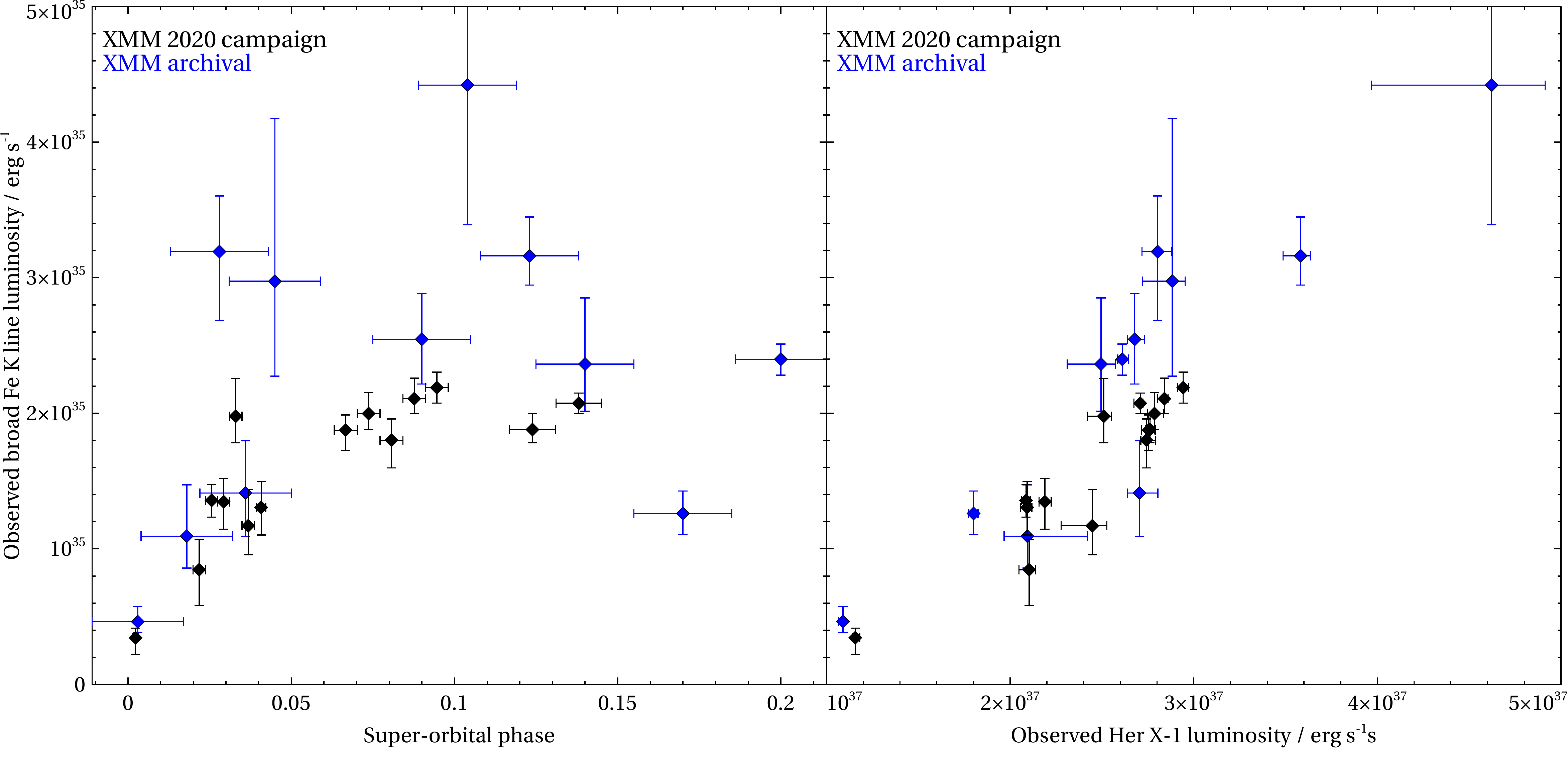}
\caption{The observed luminosity of the Broad Fe K line with respect to the super-orbital phase (left subplot) and the observed Her X-1 luminosity (right subplot). The observations with \xmm\ during the 2020 campaign, or previously are shown in distinct colours. The line could not be constrained in any \chandra\ observation. \label{FeK_lum}}
\end{figure*}

We track the evolution of the broadest Fe K line component (FWHM$\sim2$ keV) over 25 \xmm\ observations and observation segments. We find that the observed luminosity of the line strongly correlates with the observed Her X-1 luminosity (Fig. \ref{FeK_lum}, right subplot, Pearson coefficient 0.9 and p-value $9.3 \times10^{-10}$). A similar correlation with the observed Her X-1 luminosity was previously observed for the full Fe K complex (unresolved into separate line components) in \rxte\ data \citep{Vasco+12}. The observed strong correlation is consistent with the origin of the broad line as due to reprocessing of the primary continuum in the inner accretion flow of Her X-1. Alternatively, the line luminosity variation (particularly during the August 2020 campaign) may be explained as being correlated with the super-orbital phase early on in the cycle (phase$<0.1$), after which the luminosity becomes roughly constant (around phase$\sim0.1$ and beyond). This can be seen in Fig. \ref{FeK_lum}, left subplot).

\begin{figure}
\includegraphics[width=\columnwidth]{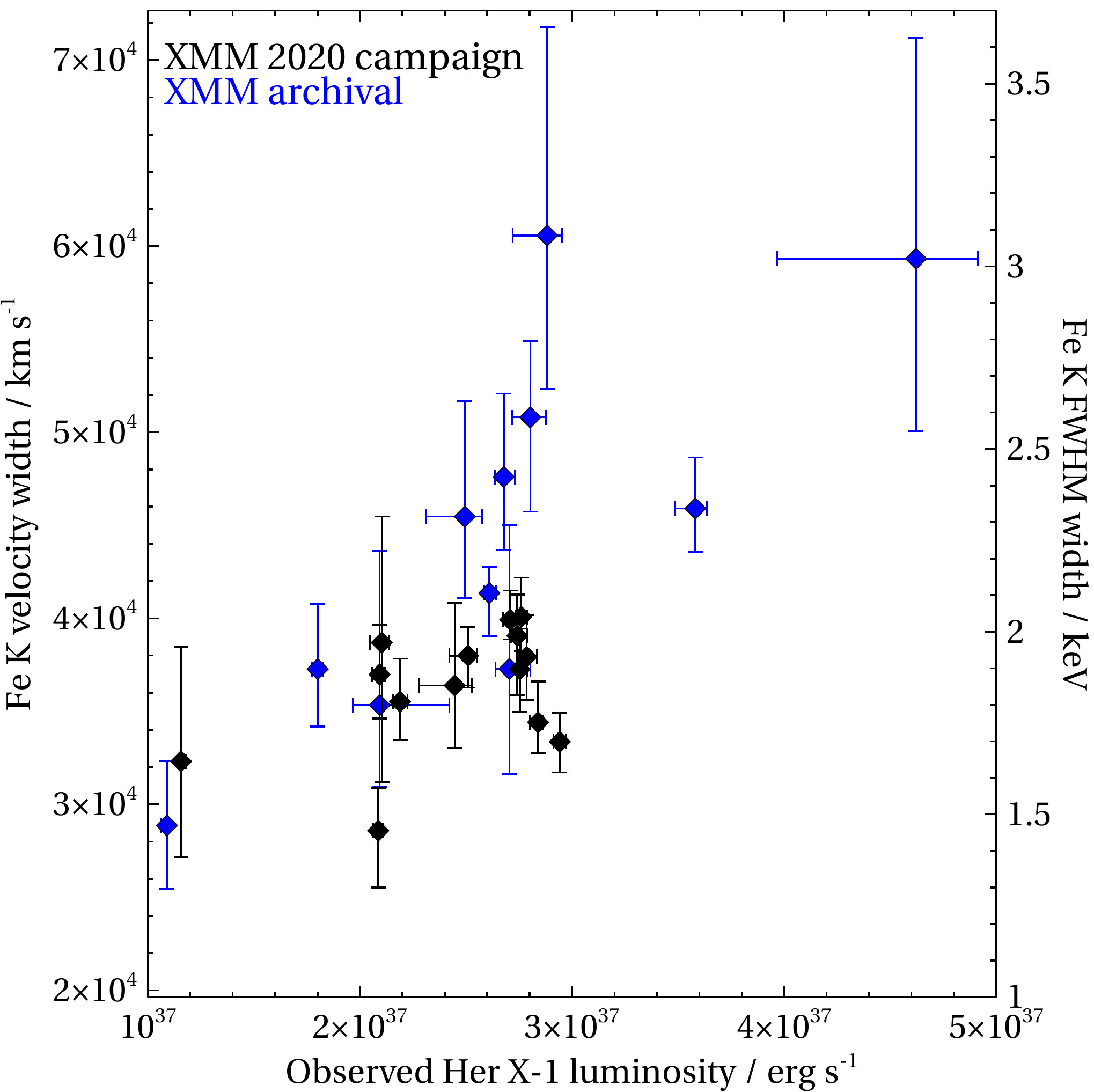}
\caption{The Broad Fe K line width versus the observed Her X-1 luminosity. The observations with \xmm\ during the 2020 campaign, or previously are shown in distinct colours. The line could not be constrained in any \chandra\ observation. \label{FeK_width}}
\end{figure}

The width of the line is extreme and requires large 1D velocities. In our observations the FWHM width (shown in Fig. \ref{FeK_width}) is inconsistent with a constant and varies from 1.5 keV to 3 keV (velocity widths from 30000 to 60000 km/s). The line is narrower when the observed Her X-1 luminosity or super-orbital phase is low. The best-fitting energy of the Fe K line is inconsistent with a constant, and varies from 6.4 keV (pegged to the lowest allowed energy during spectral fitting) up to 6.6 keV. However, no clear correlation in the evolution of the energy with the observed Her X-1 luminosity or super-orbital phase is observed. 

\begin{figure*}
\includegraphics[width=\textwidth]{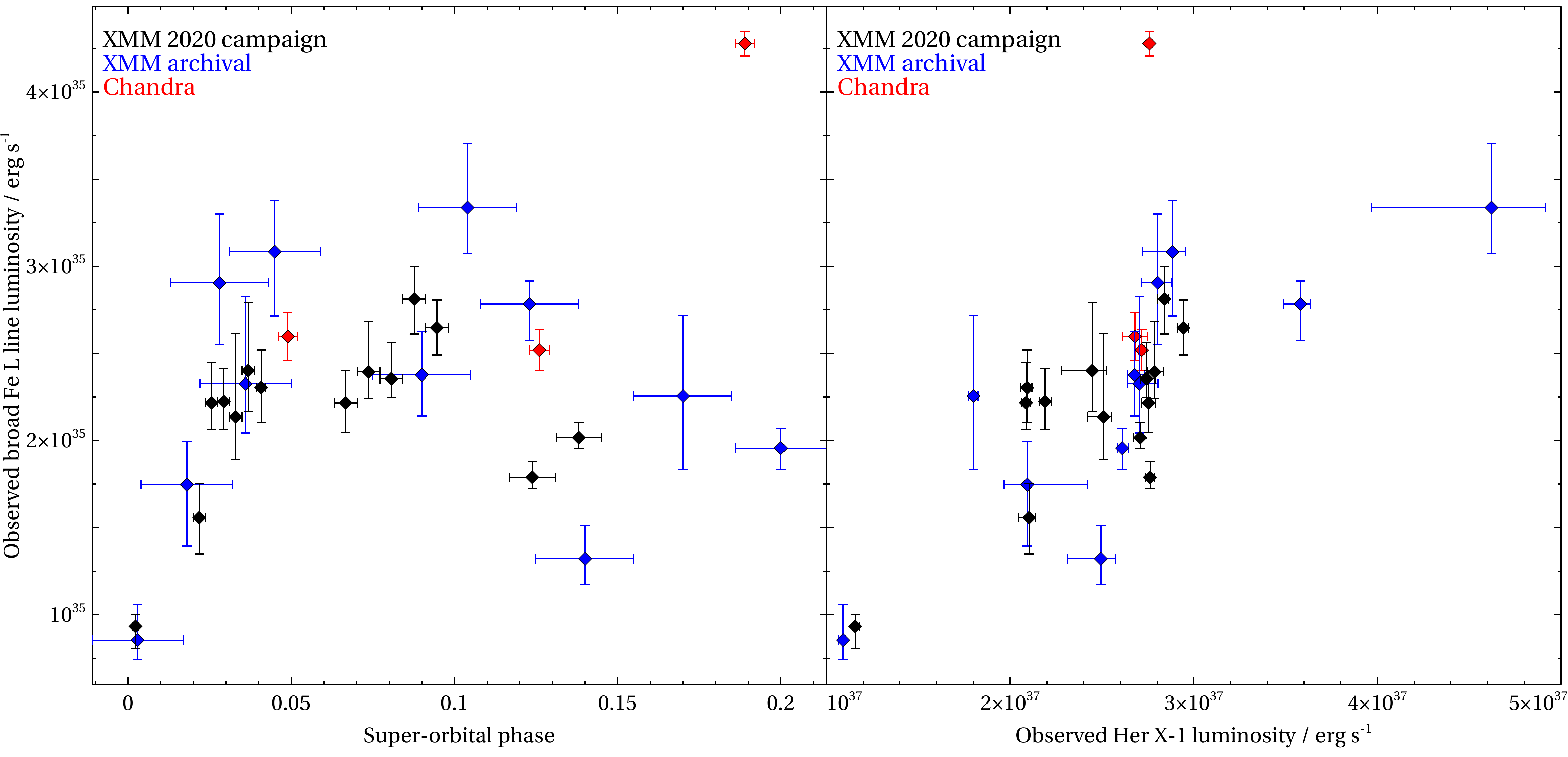}
\caption{The Broad Fe L line (1 keV residual) width versus the super-orbital phase (left subplot) and the observed Her X-1 luminosity (right subplot). The observations with \xmm\ during the 2020 campaign, previously, and with \chandra\ are shown in distinct colours. \label{FeL_lum}}
\end{figure*}

The broad Gaussian around 1 keV, possibly originating from a forest of Fe L emission lines (with likely contribution from Ne IX and Ne X), appears similarly extreme as the Fe K feature. With an energy of $0.95-0.96$ keV, it shows FWHM widths ranging from 0.32 to 0.45 keV (velocity widths from 40000 to 60000 km/s). There is no strict trend of the widths with the observed Her X-1 luminosity or super-orbital phase, but lower phases tend to show higher 1 keV line widths. The observed luminosity of this component is highly correlated with Her X-1 luminosity (Pearson coefficient 0.68, p-value $6.3\times10^{-5}$), similarly to the Fe K broad component (Fig. \ref{FeL_lum}, right subplot). There is one clear outlier, which is the archival \chandra\ observations. However, the remaining \chandra\ observations (from August 2020) seem to agree well with the \xmm\ measurements. Similarly as with the Fe K broad component, the observed luminosity evolution could instead be interpreted to evolve with the super-orbital phase as a fast rise after \textit{Turn-on} (phases $<0.05$), followed by roughly constant luminosity for the remainder of the sampled \textit{High State} (Fig. \ref{FeL_lum}, left subplot).

\section{Discussion} \label{sec:discussion}

The excellent Her X-1 \xmm\ and \chandra\ dataset, including the recent August 2020 large campaign, allows us to study the time variation of various spectral lines and components in the luminous and nearby neutron star X-ray binary Her X-1. We were able to split the available data into 28 intervals for an extensive time-resolved, high-spectral resolution study. In this paper, we focus on the evolution of the emission lines observed in the \textit{High State} of Her X-1. 

The lines we studied can be separated into three groups by velocity width: narrow lines with widths of about 1000 km/s (Fe I, O VII(i), N VI(i)), medium-width lines with widths of $5000-10000$ km/s (Fe XXV, O VIII, N VII) and extremely broad lines with velocity widths of $30000-60000$ km/s (Fe K, Fe L). Below we discuss the plausible nature of the three emission line groups, and the location of the plasma which emits them.

\subsection{Origin of the narrow lines}

The soft X-ray \textit{High State} spectra contain strong narrow intercombination lines of O VII and N VI. The presence of these transitions (and the lack of strong forbidden lines) indicates either collisional excitation due to a high density environment \citep[at least $10^{10}$ cm$^{-3}$,][]{Porquet+00}, or alternatively photoexcitation of the forbidden lines of the He-like triplets by intense UV radiation \citep{Mewe+78}. The same transitions also dominate the \textit{Low State} spectra, and were previously discussed in detail by \citet{Jimenez+02} and \citet{Jimenez+05}.

We find that the measured widths ($500-2000$ km/s) of these lines are significantly larger than the broadening found in the same transitions in previous Her X-1 studies of the Low and the \textit{Short High} states \citep[$<$300 km/s,][]{Jimenez+02, Jimenez+05, Ji+09}. The soft X-ray lines are thought to originate in the ionized atmosphere and in the accretion disk corona above the disk \citep{Jimenez+02b}. From the very low widths seen in the lower flux states, it is likely that only the very outermost part of the atmosphere/corona is observable, and the inner parts are hidden from our sight by the warped disk. Our findings in this work indicate that during the \textit{Main High} state, we are able to view the corona/atmosphere at radii closer to the central compact object, where the Doppler motions due to orbital motions are larger. Velocity widths of about 1000 km/s correspond to disk radii of $\sim2 \times 10^{10}$ cm, compared with the radii of $(8-10)\times 10^{10}$ cm estimated by \citet{Jimenez+05} from the \textit{Low State} spectra. For reference, the outer radius of the accretion disk is thought to be about $2\times10^{11}$ cm \citep{Cheng+95}.

In fact, the line broadening of the narrow lines is similar to the outflow velocity of the highly ionized disk wind \citep[$300-1000$ km/s,][]{Kosec+20}. However, we note that the ionization parameter of the wind \citep[$\log (\xi$/erg cm s$^{-1}$) of $3-4$,][Paper II]{Kosec+20} is about two orders of magnitude higher than the ionization parameter required for the production of O VII(i) and N VI(i) \citep[$\log (\xi$/erg cm s$^{-1}$) of $1-2$,][]{Ji+09}. Therefore they cannot originate from the same plasma, but they could be located at similar distances from the X-ray source. For instance, the intercombination lines could originate in high density clumps (or an accretion disk atmosphere) near the base of the disk wind.

We also detect a significant variation of Fe I line flux with super-orbital phase, indicating that most of its flux during the \textit{High State} cannot originate from the heated surface of the secondary HZ Her. A strong evolution with orbital phase was observed in the \textit{Low State} \citep{Zane+04}, but this is not the case in the \textit{High State}. Instead, most of the flux of the Fe I line likely originates from dense, neutral or very low ionization clumps in the accretion disk or the ADC at distances of several $10^{10}$ cm from the neutron star - at similar locations where O VII(i) and N VI(i) are produced. The line cannot originate very close to the neutron star given its low width ($\sim$1000 km/s at most), but most of its luminosity also cannot arise beyond the outer disk given the clear evolution with the super-orbital phase.

 The Fe I luminosity could then naturally vary with super-orbital phase as the area of the warped disk (and the ADC) observed from our point of view increases over time during the \textit{High State} phase of disk precession \citep[similar to the proposed cause of variation in the reprocessed blackbody emission,][]{brumback+21}. These regions are obscured by the warped disk in the \textit{Low State}. The Fe I material is plausibly even denser than the plasma producing the soft X-ray lines. These findings underline the strongly multiphase nature of the disk atmosphere/corona. Multiphase ADC structure in Her X-1 was previously discussed by \citet{Jimenez+05} and was also found in other ADC X-ray binaries \citep{Psaradaki+18}. 

\subsection{Origin of the medium-width lines}

The velocity widths of the three medium-width lines, Fe XXV, O VIII and N VII are close to each other and vary from $10000-15000$ km/s at the beginning of the \textit{High State} down to about $3500-4000$ km/s towards the end of the \textit{High State} (Fig. \ref{FeXXV_OVIII_width}). The similar trend of the velocity widths with super-orbital phase could indicate a common origin. Their widths correspond to circular velocities in the range between 400 and 7000 \RG. This region is close to the inner accretion disk boundary and the neutron star magnetosphere, thought to be around $2\times10^{8}$ cm from the neutron star ($\sim$1000 \RG), given the Her X-1  field of $10^{12}$ G \citep{Truemper+78}. We therefore argue that all three spectral features have a common origin in accretion flows in the vicinity of this region. The magnetosphere boundary could create a shock, or the boundary region could be strongly illuminated by the accretion column radiation, and so the plasma may be strongly ionized, reprocessing the primary radiation and producing the observed emission lines. 


However, we do not find evidence for double-peaked broad emission lines, as seen in 4U 1626-67 \citep{Schulz+01, Hemphill+21}. Such a shape might be difficult to identify for the Fe XXV line (located in the complex Fe K region), but can be clearly rejected for the O VIII and N VII lines, observed at higher spectral resolution. The accretion flow in Her X-1 is oriented almost edge-on towards the observer, and therefore the red and blue components should be easily separable using the RGS spectra (given the expected velocities of 1000s km/s), if the inner disk edge is fully visible at each super-orbital phase. If only a segment of the inner edge is visible (due to obscuration), the medium-width lines might not have a standard double-peaked diskline shape, and could still originate from X-ray reflection of the primary continuum. Over the super-orbital cycle, the visible disk segment may change, potentially explaining the observed systematic shifts in the line centroids (Fig. \ref{OVIII_NVII_wave}). Alternatively, the emission lines could originate in a shock around the magnetospheric boundary. In this case the expected line shape is difficult to predict. Further work is required to determine the exact emission process.


We note that similar medium-width emission lines ($2000-10000$ km/s widths, particularly O VIII and N VII) have been detected in some ultraluminous X-ray sources \citep[e.g.][]{Pinto+20, Pinto+21}. At least a fraction, but possibly a majority of ULXs are powered by magnetized neutron stars, accreting highly above their Eddington limit \citep[e.g.][]{Bachetti+14, Fuerst+16, Israel+17, Koliopanos+17, Walton+18}. Furthermore, broadened emission lines of O and N were recently also detected in the highly magnetized neutron stars SMC X-3 and RX J0209.6-7427, accreting mildly in excess of their Eddington limits \citep{Koliopanos+18, Kosec+21}. All these features, detected in a range of objects could originate in equivalent physical regions in the vicinity of the magnetospheric boundary. If this is indeed the case, the width of the detected features could be used to estimate the magnetosphere size (and the surface magnetic field strength) for each system. Further work is necessary to establish or reject this apparent similarity between the different accreting systems.

\begin{figure}
\vspace{0.2cm}
\includegraphics[width=\columnwidth]{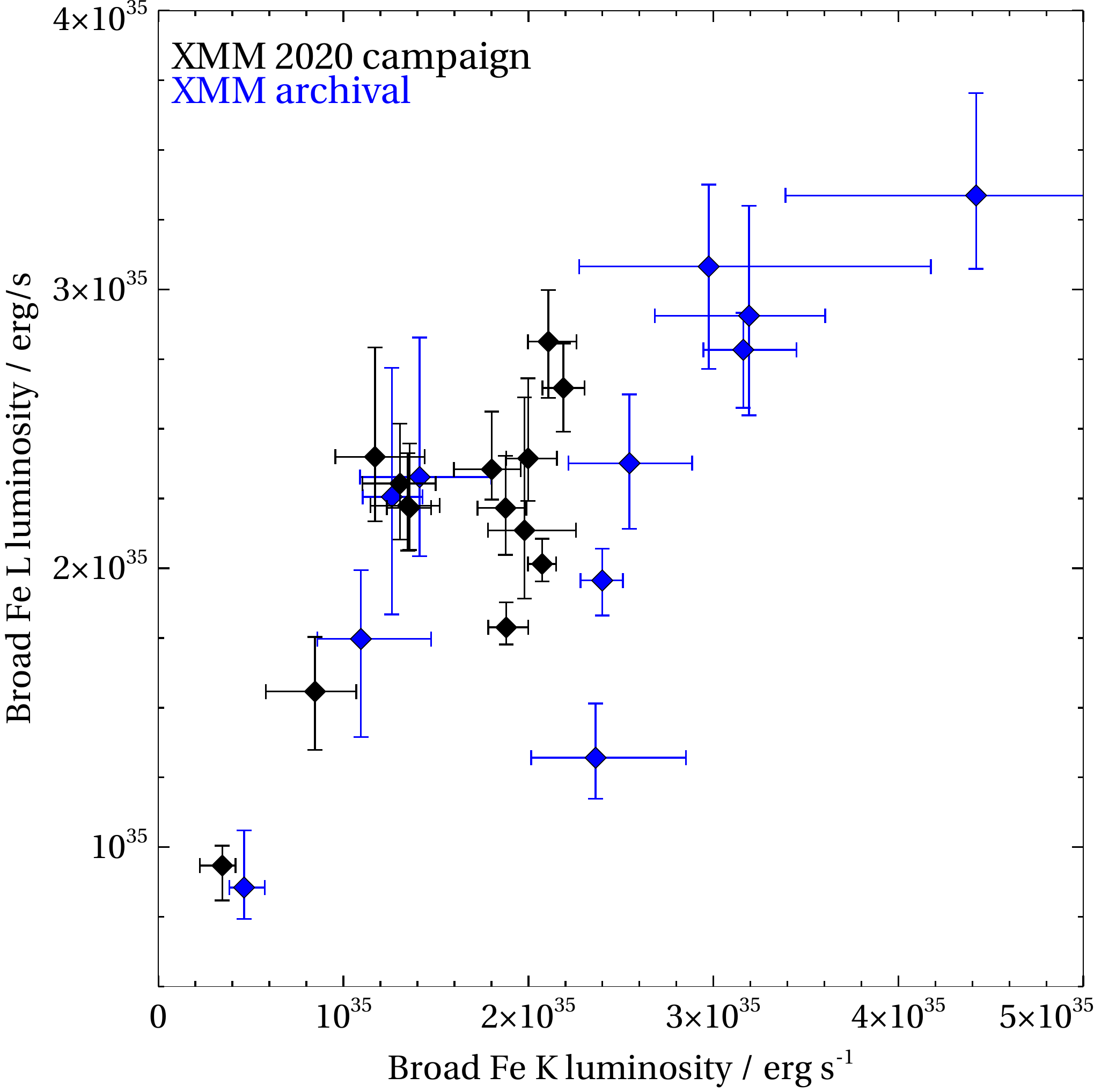}
\caption{The observed luminosity of the broad Fe L line versus the observed luminosity of the broad Fe K line. The observations with \xmm\ during the 2020 campaign, and previous observations are shown in distinct colours. \label{FeK_vs_FeL}}
\end{figure}

\subsection{Origin of the broad lines}

\citet{Asami+14} previously discussed the possible nature of the broad Fe K feature. They rejected the origin as a high column density partial covering (neutral or ionized) absorber, or line broadening due to Comptonization from the accretion disk corona. They found no plausible line origins, given the likely disk truncation radius of $\sim1000$ \RG\ (with maximum orbital velocities of 10000 km/s).

We find that the velocity widths of the broad Fe L and Fe K components are largely similar (30000 to 60000 km/s, 0.1 to 0.2c). Furthermore, both components are highly correlated with the observed Her X-1 luminosity and the correlations are even tighter with the observed primary Comptonization luminosity (Pearson coefficients 0.92 and 0.69 for Fe K and L, respectively). As a result the apparent luminosity of both lines is clearly correlated (Pearson coefficient 0.75, p-value $1.9\times10^{-5}$), which is shown in Fig. \ref{FeK_vs_FeL}. These findings likely indicate a common origin of the two broadest features in reprocessing \citep[X-ray reflection,][]{George+91, Ross+05} of the primary continuum.

However, considering that the disk is most likely truncated by the magnetosphere at $\sim1000$ \RG, it is impossible that such a broad X-ray reflection component could originate in reprocessing of primary continuum by the inner accretion disk. The disk would have to extend in as close as 25  \RG\ to the neutron star to reach orbital velocities as large as the observed line velocity widths. If these broad lines arise in a standard relativistic X-ray reflection off of the inner accretion disk as seen in other NS XRBs \citep{Ludlam+17}, and black hole XRBs \citep{Remillard+06}, then an accretion disk would have to exist within the neutron star magnetosphere, or the magnetosphere would have to be much smaller (e.g. if the magnetic field is multi-polar). The broad lines also cannot originate from the neutron star surface considering the line shapes are at least roughly symmetric - it is not clear how the blueshifted line wing could arise from the (almost stationary) highly gravitationally redshifted surface. 

The only other possibility is the reprocessing of the primary accretion column radiation by the column/accretion curtain itself. The matter infalling along the neutron star field lines has a low toroidal velocity given the slow rotation frequency of Her X-1, but it can easily reach velocities above 0.1c in the direction of the field lines before reaching the star surface. At different neutron star rotation phases, the projected velocity vector of the matter following the field lines can therefore produce both significant blueshifts and redshifts.

The accretion column and curtain are therefore the best candidates for the origin of these very broad emission features. The accretion curtain, or part of it could be optically thick \citep{Mushtukov+18} and thus plausibly produce an X-ray reflection spectrum. If this interpretation is correct, it represents the first detection of an X-ray reflection signature from the magnetically channeled flow of a neutron star. 



A broad Fe K line (up to 2 keV FWHM) was detected in the transient super-Eddington neutron star X-ray binary Swift J0243.6+6124 \citep{Jaisawal+19}. A cyclotron scattering feature was not found in its X-ray spectrum and so its magnetic field (and magnetosphere size) cannot be directly measured as in the case of Her X-1. \citet{Jaisawal+19} concluded that the broad line could thus be either due to X-ray reprocessing from the inner disk (requiring a very low magnetic field of $10^{11}$ G), from the accretion curtain, or from a photoionized ultrafast outflow.

In our further work, we will apply X-ray reflection models to describe the Her X-1 broad line emission physically and study the properties of the accretion curtain - its geometry, and possibly the density of the reprocessing layer.

We further note that reprocessing of the primary continuum by an accretion curtain will not produce a standard relativistic X-ray reflection spectral shape with a sharp blue wing and a broad red wing \citep{Fabian+89, George+91}. First, the accretion curtain geometry is significantly different from that of a standard thin disk. Secondly, and more importantly, the whole curtain structure rotates with the rotation period of the neutron star. The periodic rotation could qualitatively explain the roughly Gaussian shape of the emission lines, observed instead of a strongly double-peaked, disk line shape.

Importantly, if the broad Fe features originate in the accretion curtain, they will show periodic variations over the neutron star pulse period. Tentative variability in the pulse-resolved broad Fe K line energy was previously detected by \citet{Fuerst+13} using \nustar\ data (Fig. 9 of their paper). The Fe K line normalization is also variable over the pulse phase \citep{Vasco+13}. We will test this hypothesis with a pulse-resolved spectral analysis of the August 2020 \xmm\ dataset, in which such variation could be detected in more detail at a much higher statistical significance (thanks to the superior spectral resolution of EPIC pn). This will be addressed in a future publication.

\section{Conclusions} \label{sec:conclusions}

To sum up, our analysis of the extensive \xmm\ and \chandra\ observations suggests that various emission lines detected in the \textit{High State} Her X-1 X-ray spectrum originate in physically very distinct regions of the accretion flow. Our interpretation of their origin is summarized in a schematic in Fig. \ref{HerX1_scheme}. Our conclusions are as follows:

\begin{itemize}
    \item The narrow emission lines of Fe I, O VII(i) and N VI(i), with velocity widths of $500-2000$ km/s likely originate in the atmosphere (ADC) above the outer parts of the warped accretion disk, roughly $10^{5}$ \RG\ away from the neutron star. These lines show significantly larger widths than the same transitions observed during the \textit{Low State}, indicating that during the \textit{High State} we are able to observe the inner parts of the atmosphere closer to the neutron star, while only its outskirts (at $10^6$ \RG) are observable in the \textit{Low State}. Such findings agree with the precessing warped disk interpretation of the long-term Her X-1 flux variability.
    
    \item The medium-width emission lines of Fe XXV, O VIII and N VII, with velocity widths of $3000-10000$ km/s likely also share a common origin. They may originate at the boundary between the inner accretion disk and the magnetosphere of the neutron star, about $10^{3}$ \RG\ away from the compact object.
    
    \item The broad emission lines, denoted Fe K and Fe L, show widths of $0.1-0.2$c, and must originate closer to the neutron star than the inner accretion disk edge. Their luminosities strongly correlate with the observed luminosity of Her X-1, and hence they are most likely produced by reprocessing of the primary continuum by the accretion curtain of the neutron star, at distances of $\sim10^2$ \RG\ from the star. The relativistic motion of the accretion curtain material, infalling along the rotating magnetic field lines naturally imprints the observed extreme line widths.
\end{itemize}

\begin{acknowledgments}
 Support for this work was provided by the National Aeronautics and Space Administration through the Smithsonian Astrophysical Observatory (SAO) contract SV3-73016 to MIT for Support of the Chandra X-Ray Center and Science Instruments. PK and EK acknowledge support from NASA grants 80NSSC21K0872 and DD0-21125X. RB acknowledges support by NASA under Award Number 80GSFC21M0002. CRC acknowledges support by NASA through the Smithsonian Astrophysical Observatory (SAO) Grant GO8-19011C to MIT, and by contract SV3-73016 to MIT for Support of the Chandra X-Ray Center (CXC) and Science Instruments. This work is based on observations obtained with \xmm, an ESA science mission funded by ESA Member States and USA (NASA).
\end{acknowledgments}

%

\facilities{\xmm, \chandra, \swift, \maxi}


\software{SPEX \citep{Kaastra+96}, XSPEC \citep{Arnaud+96}, Veusz}



\appendix

\section{The gain shift of the EPIC-pn Timing mode spectra}
\label{app:pngain}

\begin{figure}
\includegraphics[width=\textwidth]{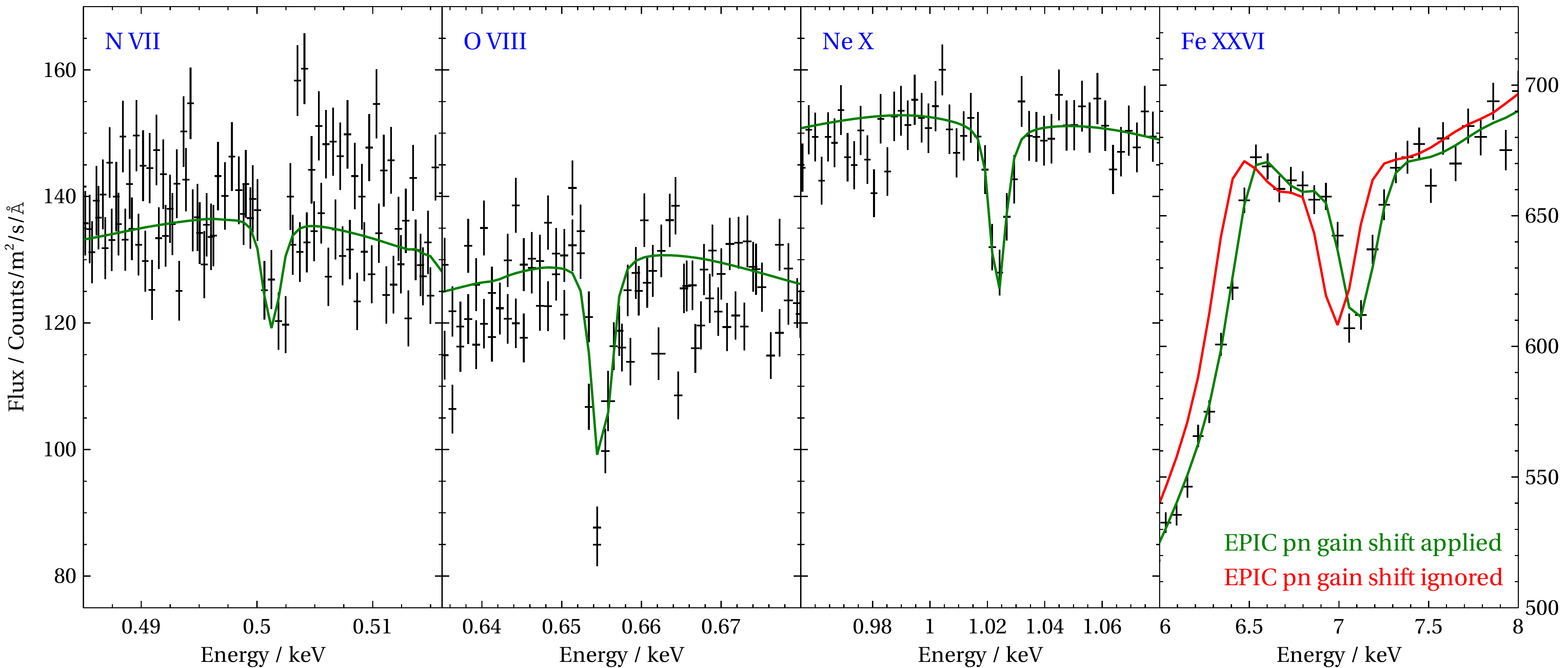}
\caption{\xmm\ RGS 1, RGS 2 and EPIC pn spectra from the \textit{High State} part of observation 0865440101. Only narrow energy bands around the strongest disk wind absorption lines N VII, O VIII, Ne X and Fe XXVI are shown. The first three subplots contain RGS 1 and RGS 2 data only, the fourth subplot only contains EPIC pn data. The fourth subplot shows the effect of applying or removing the EPIC pn gainshift correction, described in Appendix \ref{app:pngain}. The Fe XXVI residual can be correctly aligned with the residuals in RGS spectra by applying a blueshift to the EPIC spectral model of about -0.016c (a linear shift of 110 eV at 7 keV). \label{gain_abslines}}
\end{figure}

During the combined \xmm\ RGS and EPIC pn spectral analysis, we found a problem with the energy gain of the EPIC pn (Timing mode) spectra. In general, we find that the apparent photon energies in EPIC are too large, by as much as 100 eV at $6-7$ keV. The issue was initially discovered when fitting the ionized wind absorption lines. While the absorption lines within the RGS band aligned themselves perfectly, the Fe XXV/XXVI lines in the EPIC energy band always appeared offset. This is shown in Fig. \ref{gain_abslines}, where the photoionization model is centred on the N VII, O VIII and Ne X lines in the RGS band but is clearly offset for the Fe XXV/XXVI absorption. 

Additionally, we found that while fitting the spectra of the absorption dips of Her X-1, the Fe K edge at 7.1 keV was always systematically offset, apparently blueshifted by $\sim$0.01c. Furthermore, the best-fitting energy of the prominent (in dip spectra) narrow Fe I line was often around $6.45-6.48$ keV in the EPIC pn fits. Suspiciously, the Fe I line was always very close to 6.40 keV in all \chandra\ HETG spectra of Her X-1.

The findings above suggest a gain inaccuracy in the EPIC pn Timing mode. A similar gain issue, but in other EPIC pn observing modes was previously discussed by multiple authors, including \citet{Ponti+13}, \citet{Cappi+16}, \citet{Zoghbi+19}, \citet{Grafton-Waters+21} and \citet{Costanzo+21} but with a significantly lower value of about 50 eV (in the same `blueshift' direction as found here). The shift could be caused by uncorrected charge transfer inefficiency (CTI) evolution in EPIC pn or alternatively by optical loading, and appears to be significantly worse in Timing mode spectra.

A gain shift of the order of 100 eV at $6-7$ keV is detrimental to our ionized absorption analysis as well as to our study of the Fe K emission lines. This is due to the excellent counting statistics of our EPIC spectra which can easily drive the spectral fits into completely wrong parameter spaces if the data are not calibrated precisely. The shift therefore had to be corrected, but it was not clear what the shift value is or whether it is constant at all, even within the August 2020 \xmm\ campaign. We found that the gain shift during the absorption dipping states (where overall count rates are lower) appeared lower than during the full \textit{High State} observations with extreme count rates.

\begin{figure}
\includegraphics[width=\textwidth]{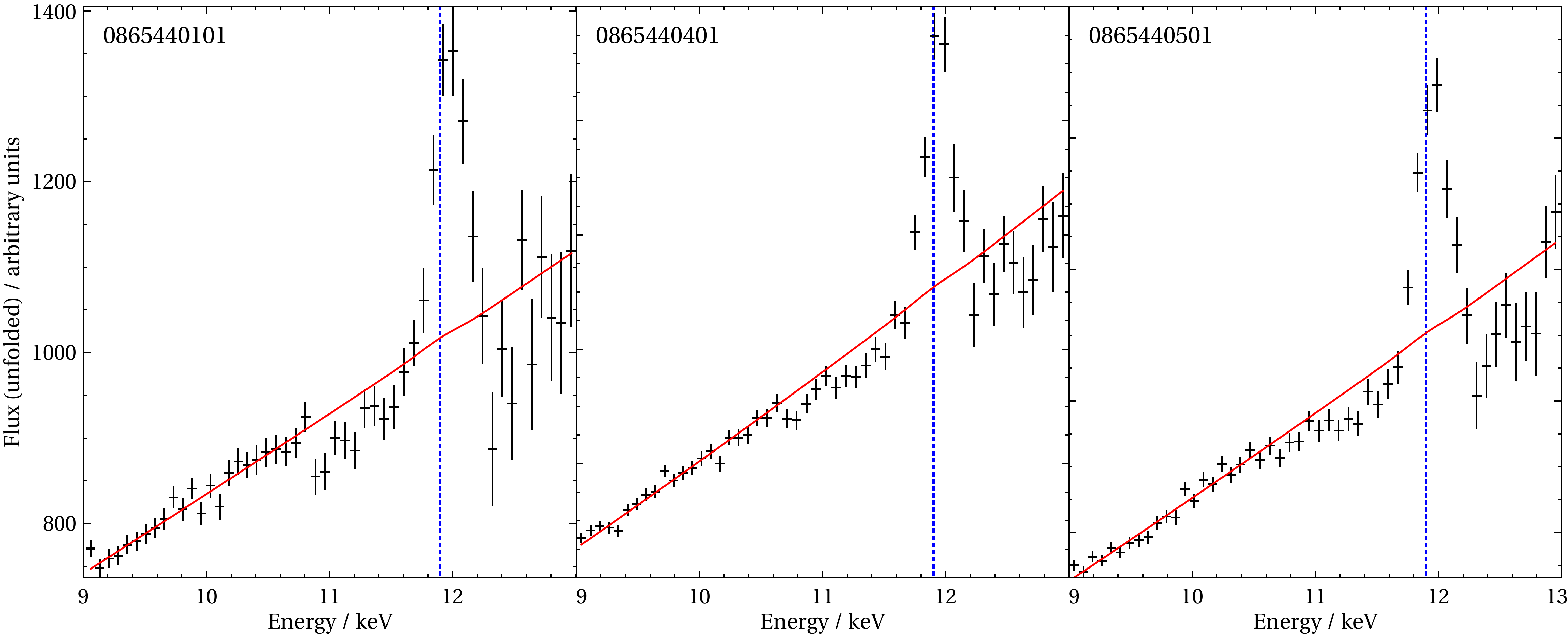}
\caption{EPIC pn $9-13$ keV spectra of the Her X-1 High flux states from the three August 2020 \xmm\ observations. The spectra are fitted with simple powerlaw models, and all show strong residuals around the instrumental Au M-edge (11.9 keV, blue dotted line), indicating a significant gain shift at these energies. \label{gain_AuMedge}}
\end{figure}

If the gain is shifted, the incorrectly assigned energies of photons around the instrumental edges such as Si K-edge (1.8 keV), Au L-edge (2.2 keV) and Au M-edge (11.9 keV) should create residuals in the source spectra. We indeed observe very strong residuals exactly at the Au M-edge, shown in Fig. \ref{gain_AuMedge}, in all three   August 2020 \xmm\ observations. The residuals can be removed from the spectra by shifting the photon energies by $100-200$ eV (exact value depends on the observation).

We first attempted to correct the issue by using the \textsc{gain} function in \textsc{xspec}, which allows to shift the energy scale of a spectrum with a multiplicative and a linear parameter. We used the residuals around the positions of the instrumental edges and fitted for the gain shift using both multiplicative and linear parameters, following the approach described in  XMM-CAL-SRN-0369\footnote{xmmweb.esac.esa.int/docs/documents/CAL-SRN-0369-0-0.pdf}. Unfortunately, this did not produce satisfactory results with a significant gain shift variation and the shift values were changing depending on the spectral model chosen (both a multiplicative and additive shift was required). The results of the gain fitting using this method could be affected by residual pile-up above 10 keV (which we did not correct for) as well as residual pile-up around 2 keV. The residual pile-up effects around 2 keV should not be significant but the count rate at these energies is very high, leading to extremely small errorbars. Finally, there are possibly real, physical emission features from Si and S in this energy band. As a result, any inaccuracies made in the 2 keV band (further compounded by the limited EPIC spectral resolution at these energies) can have detrimental effects on our ability to correct the gain shift in the Fe K band at $6-7$ keV.

Therefore, due to the issues described above, as well as due to issues with exporting our photoionization model \textsc{pion} into \textsc{xspec}, we decided to simplify our gain correction approach and perform it completely within \textsc{spex}. Our method is similar to the gain shift correction of the EPIC pn Timing mode observations of Cyg X-1 by \citet{Duro+16}. We added a simple Doppler shift component (\textsc{reds} in \textsc{spex}) to the full spectral model, . The value of the shift (in the units of speed of light, $c$) for RGS 1 and RGS 2 spectra was fixed to 0, but was a free parameter for EPIC pn spectra (in the simultaneous RGS/EPIC fit) and the parameter was explicitly fitted for. This way we did not fit for the position of the instrumental edges as we did using the \textsc{gain} function in \textsc{xspec}, since our gainshift correction was part of the final spectral model. Therefore the EPIC spectrum had to be anchored by certain spectral model parameters. We chose to freeze the energies of the Fe I and Fe XXV lines to 6.4 and 6.67 keV, respectively, as they appeared stable in all \chandra\ HETG observations of Her X-1. Furthermore, the blueshift of the photoionization \textsc{pion} grids was always anchored by the strong narrow absorption lines in the RGS spectra (N VII, O VIII and Ne X), thus fixing the positions of the Fe XXV and XXVI absorption lines, and anchoring the EPIC spectrum with respect to the RGS spectra. We accounted for the gain shift using only one multiplicative parameter (ignoring any additive gain shifts), but the best-fitting solution was centred on the crucial Fe K band where many of our spectral components have discrete line features. Any inaccuracies outside of the Fe K band should not be critical to our modelling (our EPIC pn model outside the Fe K band is just a broad Comptonization shape).

\begin{figure}
\includegraphics[width=\textwidth]{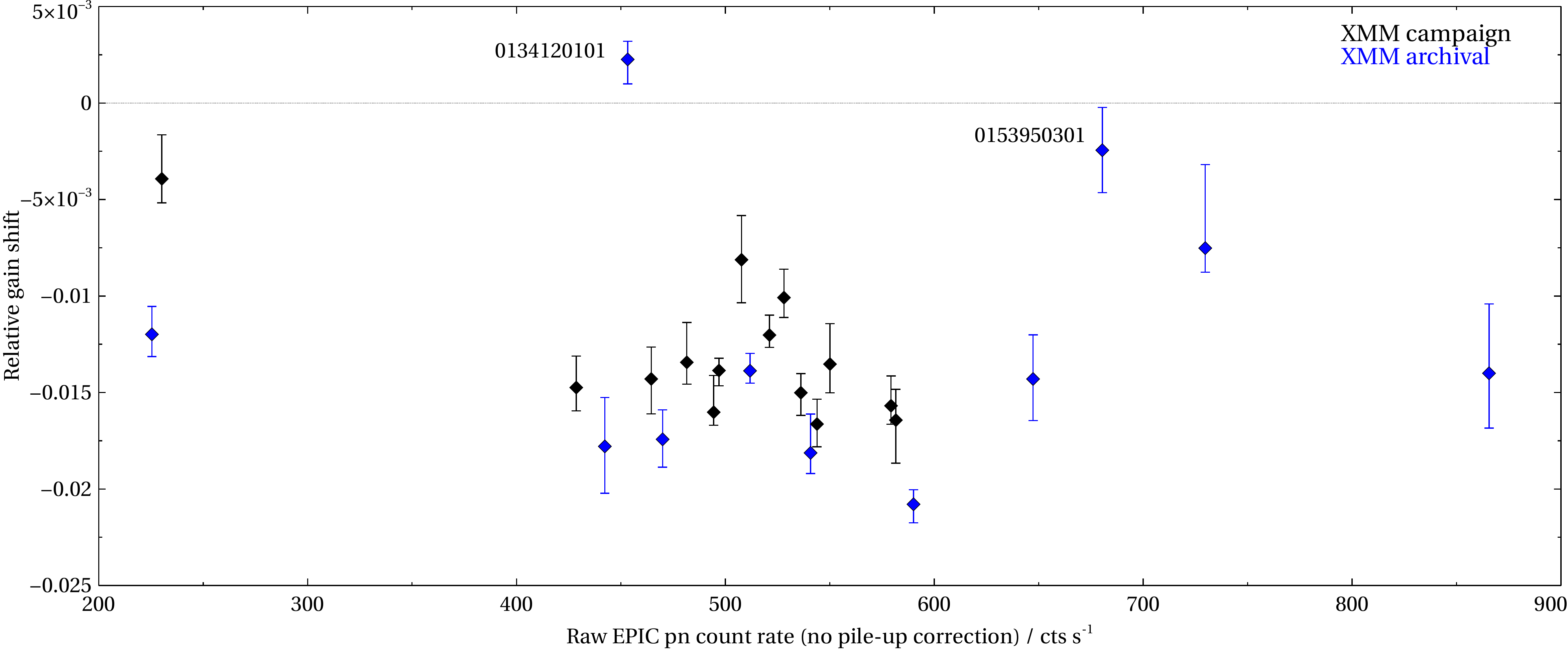}
\caption{Gainshift value (as a fraction of the speed of light) for each observation or observation segment versus raw EPIC pn count rate (no pile-up corrections applied). The August 2020 \xmm\ observations are shown in black, older archival observations are in blue. We highlight the two oldest observations carried out in 2001 and 2002. \label{gain_shift}}
\end{figure}

We track the best-fitting gainshift parameter for all the \textit{High State} observations or observational segments, but we do not find an obvious correlation between the parameter and other observational properties such as EPIC pn count rate (pile-up corrected or uncorrected), the observed Her X-1 luminosity or super-orbital phase. In Fig. \ref{gain_shift}, we show the gainshift value versus the raw pn count rate (uncorrected for pile-up). In general, the shift appears to vary between -0.01c and -0.02c for most observations but with a couple of outliers. The lower luminosity observations (0783770501 and the first segment of 0865440101) appear to need smaller shifts. We also note that the two oldest observations, 0134120101 and 0153950301, from 2001 and 2002 do not require large shifts despite high EPIC pn count rates. The remaining observations or segments do not show any strong trends. Therefore, it is unclear what is the main driver for the gain variation.

\section{Instrumental residuals in RGS or interstellar gas/dust?}
\label{app:ignrgs}

\begin{figure}
\includegraphics[width=\textwidth]{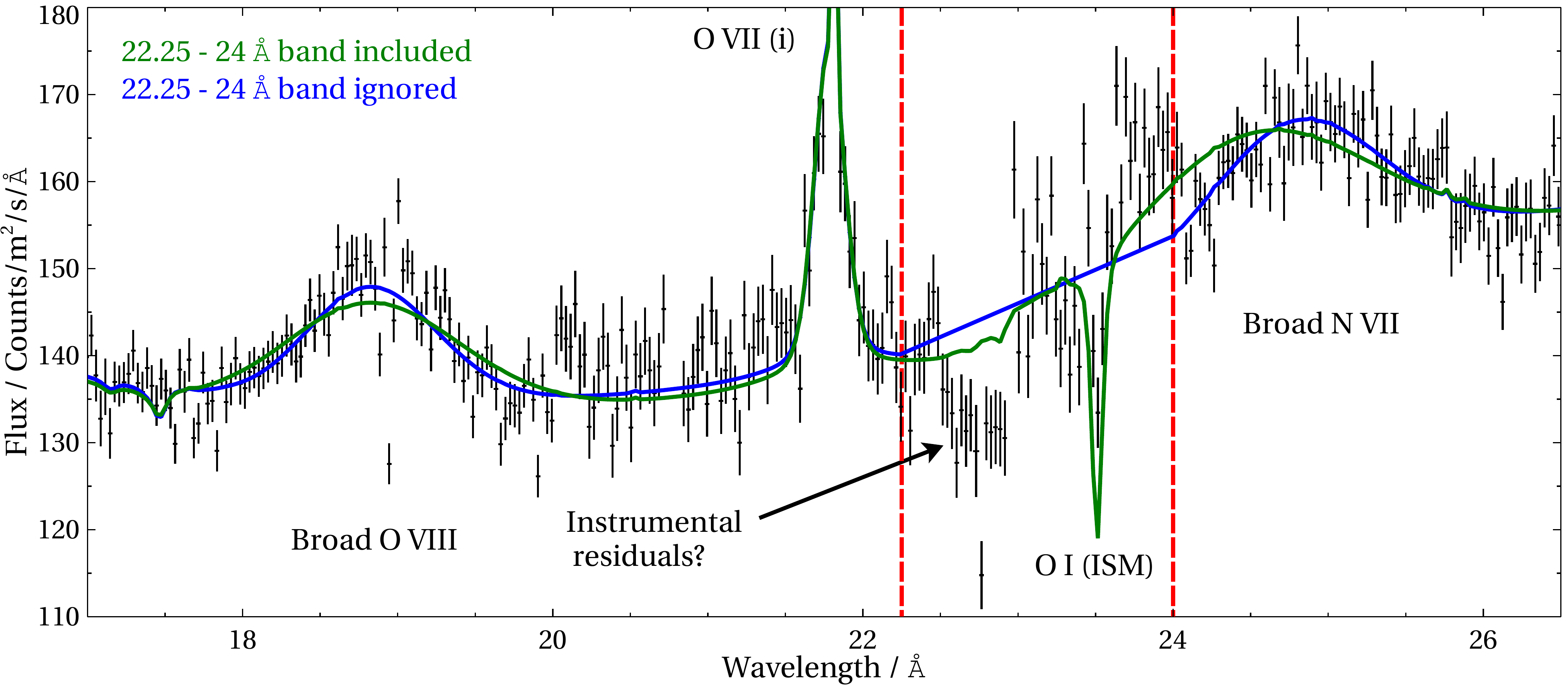}
\caption{RGS spectra (stacked for visual purposes) from observation 0865440501 around the excluded 22.25-24 \AA\ region. Two spectral models are shown and the most important model features are denoted with labels. The green model includes the problematic region while the blue model ignores it. The green model clearly over-predicts the broadening of the medium-width O VIII and N VII features. The position of the possible instrumental feature is indicated by the arrow. \label{23Aresiduals}}
\end{figure}

We further found strong residuals in the $22.5-23.5$ \AA\ range in the RGS spectra. Thanks to the excellent data quality, they are highly statistically significant. The residuals are located near the O K edge, where the absorption lines of various ionization levels of O as well as ISM dust are commonly found. We attempted to fit the features with a warm ISM plasma model as well as with the \textsc{amol} ISM dust model \citep{Pinto+10}. However, we were not able to reach good fits. A very high quality spectrum of this region (from the fully stacked observation 0865440501, with roughly 80 ks of clean exposure) is shown in Fig. \ref{23Aresiduals}. As is seen from the plot, the residuals do not appear as a set of narrow absorption features that could be fitted with a range of O lines of different ionizations. Furthermore, we note that the neutral absorption column towards Her X-1 is very low, around $1 \times 10^{20}$ \pcm\ hence no very strong warm ISM O lines are expected. Instead, what we observe is a broad absorption feature around $\sim22.75$ \AA, with a possible emission residual above 23.5 \AA. These residuals broaden our phenomenological models of the O VIII and N VII lines. Particularly the N VII feature often attempts to describe the residuals just below 24 \AA, resulting in a mis-fit of both O VIII and N VII features (considering their widths are coupled). This has detrimental effect on our ionized absorption fits in the later parts of the super-orbital cycle, when the narrow disk wind lines are very weak. 


Residuals in this energy range were present in multiple previous RGS studies of the O K edge \citep[e.g.,][]{Costantini+12,Pinto+13}. They were later discussed in particular by \citet[][Fig. 6]{Psaradaki+20}, who interpret them as instrumental features in RGS. Regardless of the origin of the features, we decided to ignore the $22.25-24$ \AA\ band from this study. The band is unimportant for disk wind modelling considering there are no strong highly ionized transitions between 22.5 and 24 \AA. It is also unlikely to have a strong impact on the modelling of the remaining included (phenomenological) emission components.

\section{Summary of spectral fit components and parameters}
\label{app:fitpar}

\begin{deluxetable*}{ccccc}
\tablecaption{Summary of the \xmm\ spectral fit components and parameters. All line widths are defined as FWHM.\label{paramtable}}
\tablewidth{0pt}
\tablehead{
\colhead{Component} & \colhead{\textsc{spex} name} & \multicolumn2c{Parameter} & \colhead{Notes}   }
\startdata
\multicolumn{5}{c}{Multiplicative components}\\
\hline
EPIC pn gain correction & \textsc{reds} & z & Gain shift parameter & \\
Interstellar absorption  & \textsc{hot} & nh & Column density & \\
Disk wind absorption  & \textsc{pion} & nh & Column density & \\
 & & $\log \xi$ & Ionization parameter &  \\
 & & v & Velocity width &  \\
 & & z & Outflow velocity &  \\
\hline
\multicolumn{5}{c}{Continuum components}\\
\hline
Primary Comptonization & \textsc{comt} & norm & Normalization &  \\
 &  & $T_0$ & Seed temperature  &  \\
 &  & $T_1$ & Electron temperature  & \\
 &  & $\tau$ & Optical depth \\
Hotter blackbody & \textsc{mbb} & norm & Normalization  & \\
 &  & T & Temperature \\
Cooler blackbody & \textsc{bb}  & norm & Normalization  &  \\
 &  & T & Temperature & fixed to 0.05 keV \\
\hline
\multicolumn{5}{c}{Broad emission lines}\\
\hline
Fe K  & \textsc{gaus} & norm & Normalization  & \\
  &  & e & Energy &  lower limit of 6.4 keV \\
  &  & fwhm &  Line width  & \\
Fe L  & \textsc{gaus} & norm & Normalization  &  \\
  &  & e &  Energy &  \\
  &  & fwhm &  Line width &  \\
\hline
\multicolumn{5}{c}{Medium-width emission lines}\\
\hline
Fe XXV  & \textsc{gaus} & norm & Normalization  &  \\
  &  & e & Energy  & fixed to 6.67 keV \\
  &  &  fwhm & Line width  & \\
O VIII  & \textsc{gaus} & norm & Normalization &   \\
  &  & w & Wavelength &  \\
  &  & awid & Line width  & \\
N VII  & \textsc{gaus} & norm & Normalization  &  \\
  &  & w & Wavelength  & \\
  &  & awid & Line width  & velocity width coupled to O VIII \\
\hline
\multicolumn{5}{c}{Narrow emission lines}\\
\hline
Fe I  & \textsc{gaus} & norm & Normalization  &  \\
  &  & e &  Energy &  fixed to 6.4 keV \\
  &  & fwhm & Line width &  fixed to 0.05 keV \\  
O VII(i)  & \textsc{gaus} & norm & Normalization  &  \\
  &  & w & Wavelength  & \\
  &  & awid & Line width  & upper limit of 0.5 \AA \\
N VI(i)& \textsc{gaus} & norm & Normalization  &  \\
  &  & w & Wavelength  & \\
  &  & awid & Line width  & upper limit of 0.5 \AA \\
\enddata
\end{deluxetable*}

\bibliography{References}{}
\bibliographystyle{aasjournal}



\end{document}